\documentclass[12pt,preprint]{emulateapj}

\usepackage{epsfig}
\usepackage{graphicx}
\def\gta{\mathrel{\hbox{\rlap{\hbox{\lower4pt\hbox{$\sim$}}}\hbox{$>$}}}}

\shorttitle{IRAC Dark Field}
\shortauthors{Krick et al.}

\begin{document}
\newcommand\msun{\hbox{M$_{\odot}$}}
\newcommand\lsun{\hbox{L$_{\odot}$}}
\newcommand\magarc{mag arcsec$^{-2}$}
\newcommand\h{$h_{70}^{-1}$}

\bibliographystyle{apj}

\title{\bf The IRAC Dark Field; Far- Infrared to X-ray Data}

\author{J.E. Krick \altaffilmark{1}, J.A. Surace \altaffilmark{1},D. Thompson \altaffilmark{2}, M. L. N. Ashby \altaffilmark{3}, J. Hora \altaffilmark{3}, V. Gorjian \altaffilmark{4}, L. Yan \altaffilmark{1}, D.T. Frayer \altaffilmark{5}, E. Egami \altaffilmark{6}, and M. Lacy \altaffilmark{1}}
\altaffiltext{1}{Spitzer Science Center, MS 220--6,
California Institute of Technology, Jet Propulsion Laboratory,
Pasadena, CA 91125, USA}
\altaffiltext {2}{Large Binocular Telescope Observatory, University of Arizona, Tucson, AZ 85721}
\altaffiltext {3}{Harvard-Smithsonian Center for Astrophysics, 60 Garden Street, Cambridge MA 02138}
\altaffiltext {4}{Jet Propulsion Laboratory, California Institute of Technology, Pasadena, CA, 91109}
\altaffiltext {5}{NASA Herschel Science Center, California Institute of Technology, Pasadena, CA, 91109}
\altaffiltext {6}{Department of Astronomy, University of Arizona, 933 N. Cherry Avenue, Tucson, AZ 85721}
\email{jkrick@caltech.edu}

\begin{abstract} 
  We present 20 band photometry from the far-IR to X-ray in the {\it
    Spitzer} IRAC dark field.  The bias for the near-IR camera on {\it
    Spitzer} is calibrated by observing a $\sim 20$\arcmin\ diameter
  ``dark'' field near the north ecliptic pole roughly every
  two-to-three weeks throughout the mission duration of {\it Spitzer}.
  The field is unique for its extreme depth, low background, high
  quality imaging, time-series information, and accompanying
  photometry including data taken with {\it Akari}, Palomar, MMT,
  KPNO, {\it Hubble}, and {\it Chandra}.  This serendipitous survey
  contains the deepest mid-IR data taken to date.  This dataset is
  well suited for studies of intermediate redshift galaxy clusters,
  high redshift galaxies, the first generation of stars, and the
  lowest mass brown dwarfs, among others.  This paper provides a
  summary of the data characteristics and catalog generation from all
  bands collected to date as well as a discussion of photometric
  redshifts and initial and expected science results and goals.  To
  illustrate the scientific potential of this unique dataset, we also
  present here IRAC color color diagrams.

\end{abstract}


\section{Introduction} 
\label{intro}
Deep, multi-wavelength surveys are an important tool in studying the
formation and evolution of galaxies.  Infrared data is particularly
useful in these surveys.  Near-IR wavelengths provide a direct measurement of the
stellar content of galaxies in absence of a dominant AGN contribution at
redshifts as high as three.  The longer wavelengths sample emission
primarily from polycyclic aromatic hydrocarbons (PAHs), as well as
direct thermal emission from hot dust.  These are markers of
star forming galaxies and active galactic nuclei (AGN).

The dark field is a multi-wavelength survey based on observations
carried out as part of the calibration for the Infrared Array Camera
\citep[IRAC;][]{fazio2004} on the {\it Spitzer} Space Telescope
\citep{werner2004}. Rather than being designed around specific science
objectives, like many other Great Observatory surveys \citep[HDF, CDF,
SWIRE, GOODS, GEMS, COSMOS, etc.;][]{ williams1996,giacconi2001,
  lonsdale2003, giavalisco2004, rix2004, scoville2007}, the survey
area and observation strategy was selected based on the calibration
requirements for IRAC.  Observations were taken to measure the dark
current and biases of the four detector arrays (see \S \ref{irac})
near the beginning and end of each IRAC observing period (or
``campaign''), resulting in 2--5 dark field measurements every
two-to-three weeks throughout the entire duration of the mission.
Each of these observations in the dark field collects roughly two
hours of integration time total in each of the four bands in a single
$5\arcmin\times5\arcmin$ field of view.  Each new visit is observed in
a slightly different position due to spacecraft orientation and
rotation.  The center of the field is located at $17^{h}40^{m}$,
$+69^{d}$ (J2000), very near the north ecliptic pole (NEP).  Over the
course of the mission, the observations have filled in a region
roughly 20\arcmin \ in diameter.  This has created the deepest ever
mid-IR survey, exceeding the depth of the deepest planned regular {\it
  Spitzer} surveys over several times their area.  Furthermore, this
is the only field for which a 5+year baseline of mid-IR periodic
observations exists\citep[see][for more inforrmation on the
time-series data]{frost2009}.


The resulting observations are unique in
several ways.  The dark field lies in the lowest possible region of
zodiacal background, the primary contributor to the infrared
background at these wavelengths, and as such is in the region where
the greatest sensitivity can be achieved in the least amount of time.
The area was also chosen specifically to be free of bright stars and
very extended galaxies, which allows clean imaging to very great
depth.  The field lies in the northern continuous viewing zone, a 10
degree radius region centered on the NEP that is always visible to
{\it Spitzer}, which means many opportunities exist to re-observe the
field essentially continuously.  The observations are done at many
position angles (which are a function of time of observation) leading
to a more uniform, circularized, final point spread function (PSF).
Finally, because the calibration data are taken directly after
anneals, they are more free of electronic artifacts such as hot pixels
and latent images than ordinary guest observer (GO) observations.

To enhance the IRAC dark field dataset we have obtained imaging data
in 16 other bands including Palomar/MMT/KPNO u', g', r', i',
z$^\prime$, J, H, K, {\it HST} /ACS F814W, Akari $4, 11, 15, \&
18\micron$, {\it Spitzer} MIPS $24 \& 70\micron$, and {\it Chandra}
ACIS-I.  Although the entire dark field is $> 20\arcmin$ in diameter,
because of spacecraft dynamics the central $\sim 15\arcmin$ is
significantly deeper and freer of artifacts.  Therefore, it is this
area which we have matched with the additional observations. All
space-based datasets are publicly available through their respective
archives.


This paper is structured in the following manner.  In \S\ref{data} we
present the data for all 20 bands, its reduction as well as source
extraction and photometry.  Details of the composite catalog are
presented in \S\ref{catalog}.  Photometric redshifts are derived and
errors calculated in \S\ref{photz}.  Lastly, a small discussion of the
science goals are presented in \S\ref{discuss}.  Throughout this paper
we use $H_0=70$km/s/Mpc, $\Omega_M$ = 0.3, $\Omega_\Lambda$ = 0.7.
All photometry is quoted in the AB magnitude system.


\section{Data}
\label{data}

\begin{figure}
\epsscale{1.0}
\plotone{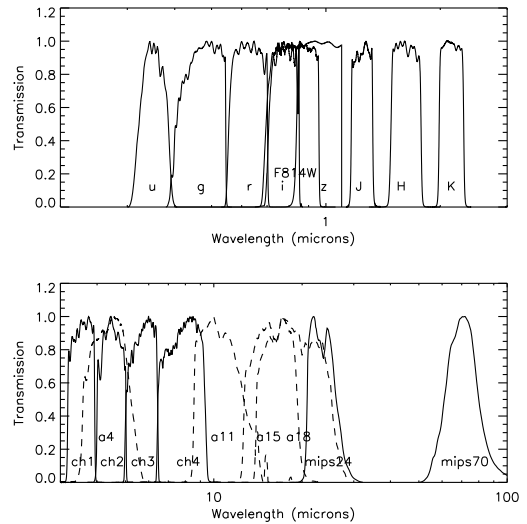}
\caption[Filter Transmission]{Filter transmission curves for the
  optical to far-IR bands.  All curves are normalized for comparison.
  The upper plot shows optical through near-IR.  Of the near-IR bands,
  only the Palomar WIRC J, H, and K curves are shown for clarity.  The
  lower plot shows the {\it Spitzer} and {\it Akari} (dashed lines)
  mid-IR to far-IR bands.}
\label{fig:filters}
\epsscale{1}
\end{figure}

\begin{figure}
\plotone{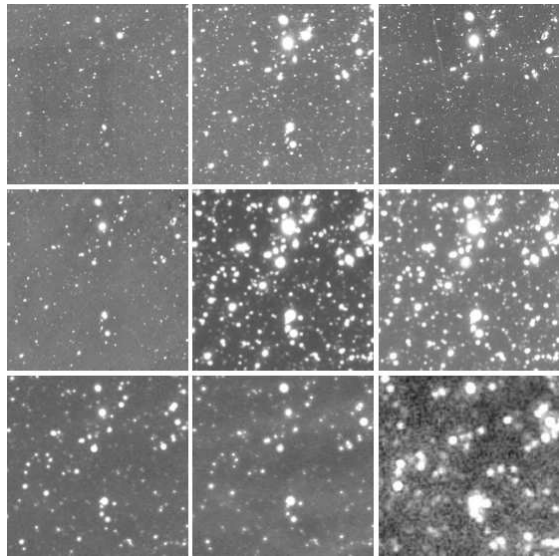}
\caption[images]{ Representative images of the field, 3 arcminutes on
  a side.  The order of the images is $u^{\prime}$, $r^{\prime}$,
  HST/ACS F814W, J, 3.6, 4.5, 5.8, 8.0, 24 \micron.}
\label{fig:images}
\epsscale{1}
\end{figure}

The details of our dataset are summarized in Table \ref{tab:data} and
in the sections below.  Data are presented roughly in order of
decreasing wavelength with the exception that the IRAC data are
discussed first.  Figure \ref{fig:filters} shows the filter curves of
the optical through infrared bands discussed below.  Figure
\ref{fig:images} shows a representative sample of data thumbnails.
\begin{deluxetable*}{lccccccc}
\tabletypesize{\footnotesize}
\tablewidth{0pc}
\tablecolumns{8}
\tablecaption{Observational Data \label{tab:data}}

\tablehead{
\colhead{Telescope} &         
\colhead{Waveband} &        
\colhead{Central$ \lambda$} &      
\colhead{Avg. Exptime \tablenotemark{a}} & 
\colhead{Area} & 
\multicolumn{2}{c}{$3\sigma$ Depth} &    
\colhead{Avg. FWHM \tablenotemark{b}}
\\
\colhead{ Instrument} &
\colhead{ } &      
\colhead{(microns)} &  
\colhead{(hours/pixel)} &
\colhead{ (\sq \arcmin)}  & 
\colhead{(AB mag.)} & 
\colhead{($\mu$Jy)} &
\colhead{(arcseconds)}
}

\startdata
   
{\it Chandra} ACIS-I         &          & 2-8 KeV  &      28.0  & 290  & $3.6\times10^{-16}$ \tablenotemark{c}&      & 5.0 \\

Palomar Hale LFC            &  u'          &   0.35  &        6.0  & 450  & 27.2                                    & 0.05 & 1.0 \\
 
Palomar Hale LFC            &  g'          &   0.47  &        6.2  & 450  & 27.1                                    & 0.05 & 1.3 \\

Palomar Hale LFC            &  r'           &   0.63  &        4.1  & 450  & 26.6                                    & 0.09 & 0.9 \\

Palomar Hale LFC            &  i'           &   0.77  &        4.7  & 450  & 26.0                                    & 0.15 & 1.2 \\

{\it HST }ACS                   &  F814W  &   0.81  &        1.4  & 260  & 28.6                                    & 0.01 & 0.1\\

MMT Megacam                &  z$^\prime$  &   0.91  &  3.6   & 580  & 25.9                                    &0 .16 & 1.0 \\

NOAO Mayall Flamingos &  J           &   1.20  &        0.5  & 220  & 21.4                                     & 10.2 & 1.2 \\

MMT SWIRC                       &  J            &   1.2  &       5.2       & 25   & 24.4                                      & 0.65 & 0.8 \\

Palomar Hale WIRC          &  J            &   1.25  &       0.8  & 120  & 21.8                                     & 7.1  & 1.1 \\

MMT SWIRC                     &  H           &   1.6 &        0.2  & 225  & 23.0                                       & 2.4 & 0.8 \\

Palomar Hale WIRC         &  H           &   1.63 &        0.7  & 120  & 22.1                                      & 5.4 & 1.1 \\

Palomar Hale WIRC         &  K           &   2.15  &        1.6  & 210  & 21.2                                     & 12.3 & 1.1 \\

{\it Spitzer} IRAC            &              &   3.6    &        8.0  & 380   & 25.7   & 0.2 \tablenotemark{d}  & 1.9 \\

{\it Spitzer} IRAC             &              &   4.5    &        10.0  & 380  & 25.9  & 0.17  \tablenotemark{d}& 1.9 \\

{\it Akari} IRC                  &              &   4.5    &        1.8  & 100  & 23.5                                      & 1.5    & 4.4 \\

{\it Spitzer} IRAC              &              &   5.8    &        8.0  & 380  & 26.3   & 0.11 \tablenotemark{d}    & 1.9 \\

{\it Spitzer} IRAC              &              &   8.0    &        10.0  & 380  & 26.3  & 0.11 \tablenotemark{d}  & 2.1 \\

{\it Akari} IRC                  &              &   11    &        2.4  & 100    & 21.0                                     & 15     & 6.2 \\

{\it Akari} IRC                  &              &   15    &        1.0  & 100   & 20.1                                       & 33    & 6.5 \\

{\it Akari} IRC                  &              &   18    &        0.8  & 100   & 19.8                                      & 45      & 6.7 \\

{\it Spitzer} MIPS              &              &   24     &        1.0  & 225  & 20.8                                      & 17.3    & 5.8 \\

{\it Spitzer} MIPS              &             &   70      &        1.6  & 160  & 14.8                                     & 3200    & 18.6 \\

\enddata
\tablenotetext{a}{Total and maximum exposure times are presented in \S \ref{data}. }
\tablenotetext{b}{Average FWHM as measured from a number of point sources in the final image. }
\tablenotetext{c}{erg/s/cm$^2$ }
\tablenotetext{d}{IRAC depths are given as measured 95\% completeness limits, see \S \ref{irac}.}

\end{deluxetable*}  


\subsection{Spitzer Space Telescope}

\subsubsection{IRAC}
\label{irac}

IRAC simultaneously observes with four detectors
at 3/6. 4/5. 5/8. \& 8/0 microns.  Each detector has a
$5.2\arcmin\times5.2\arcmin$ field of view with 1\farcs2 pixels.  Images are
acquired in pairs (3.6 and 5.8\micron, 4.5 and 8.0\micron) of two
adjacent fields, separated by 6.5\arcmin, using beam splitters. 

\begin{figure}
\plotone{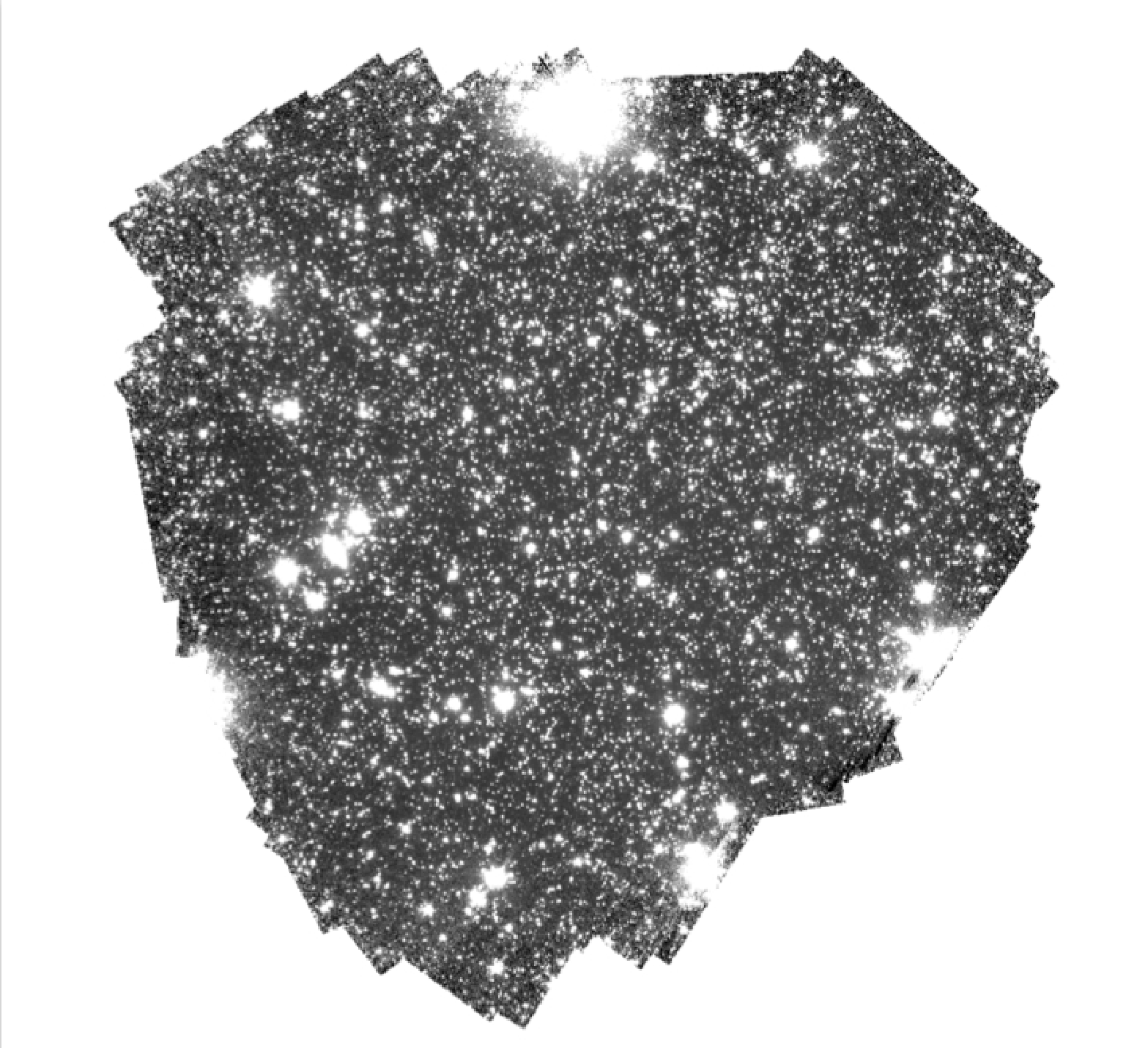}
\plotone{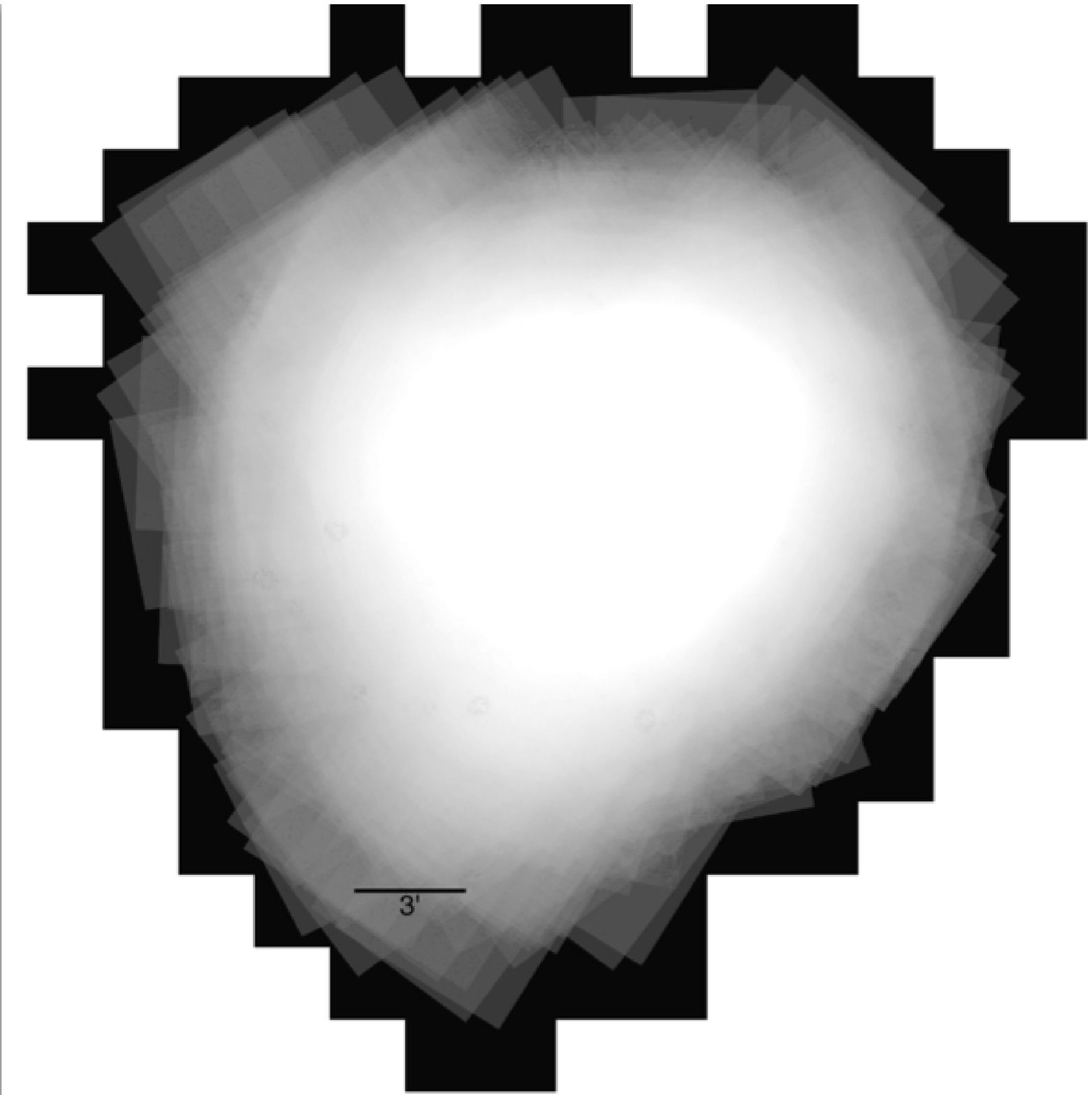}
\caption[ch1]{Top: The full 3.6 \micron\ data image.  Bottom:
  Grey-scale coverage map of the 3.6 \micron\ data with grey-scale
  ranging from 1 to 1000 images, each of 100s exposure time. North is
  up, East is to the right.  The actual NEP is 3.5 degrees away.}
\label{fig:ch1}
\epsscale{1}
\end{figure}

This work is based on a preliminary combination of calibration data
taken from Dec 2003 through Jan 2005; $\approx$30\% of the expected
dark field depth not including the warm mission, see Figure
\ref{fig:ch1}.  The 3.6, 4.5, 5.8, and 8.0 \micron\ data have a total
exposure time of 40 hours per filter covering 380 square arcminutes to
a limiting depth of 0.2, 0.17, 0.11,0.11 $\mu$Jy respectively. The
maximum per pixel exposure time is 28 hours at 3.6 and 5.8 \micron\
and 21 hours at 4.5 and 8.0\micron.  The median exposure time per
pixel is 8-10 hours; see below.  Raw astrometry from the {\em Spitzer}
pointing control system, based on inputs from the startracker and
gyroscopes, is used as the input to a pointing refinement program.
The absolute positions of stars from the Two Micron All Sky Survey
(2MASS) along with the relative positions of stars common to different
individual IRAC frames are then used to improve the astrometry from
the raw spacecraft astrometry. Final astrometric accuracy for an
individual frame is $\sim 0.3\arcsec$ rms.  Because field rotation
during the mission should average out many systematic effects, the
final astrometric accuracy of the IRAC dark field mosaic is expected
to be better than this.  The Basic Calibrated Data (BCD) products
produced by the {\it Spitzer} Science Center (pipeline versions S11,
S12, S13) were further reduced using a modified version of the
pipeline developed for the {\it Spitzer} Wide area Infrared
Extragalactic survey (SWIRE) \citep{surace2005}.  This pipeline
primarily corrects image artifacts and forces the images onto a
constant background (necessitated by the continuously changing
zodiacal background as seen from {\it Spitzer}). The data were coadded
onto a regularized 0.6\arcsec \ grid using the {\sc mopex} software
developed by the {\it Spitzer} Science Center (SSC) as described in
\citet{makovoz2005}.  Experiments we performed with DAOPHOT
demonstrate that nearly all extragalactic sources are at least
marginally resolved by IRAC, particularly at shorter wavelengths, and
hence point source fitting is inappropriate for photometry.

Photometry is done using the high spatial resolution ACS data as
priors for determining an individual, object-specific aperture shape
for extracting the IRAC fluxes.  The advantages of this are deeper
photometry and more accurate color measurements between low and high
resolution observations.  Because the IRAC beam is so large (compared
to ACS), using the higher resolution data for shape information will
allow us to deblend otherwise overlapping sources and therefore do
photometry to fainter limits (below the confusion limit) than with
IRAC alone. The disadvantage to using this technique is that we must
assume the galaxy morphology is the same across the wavelength range
of the high to low resolution data.  This assumes that ACS F814 light
is co-spatial with IR light.  At high redshifts this assumption will
be reasonable because in those galaxies we cannot resolve any age or
SF gradients across the galaxy.  For low redshift galaxies we rely on
the Rayleigh-Jeans side of the SED being less-sensitive to age so that
F814 and IRAC are measuring the same population of stars.  This
assumption is not valid for all galaxies and must be considered when
using the catalog for science goals.  The use of high resolution data
as priors for low resolution photometry has been used on similar
datasets in the literature \citep{fernandez-soto1999, papovich2001,
  shapley2005, grazian2006,laidler2007}.

This method begins by first running source detection and photometric
extraction on the coadded IRAC images using a matched filter algorithm
with image backgrounds determined using the mesh background estimator
in SExtractor \citep{bertin1996} .  This catalog is merged with the
{\it HST} / ACS catalog.  For every object in that catalog, if the
object is detected in ACS then we use the ACS shape parameters to
determine the elliptical aperture size for the IRAC images.  ACS shape
parameters are determined by SExtractor on isophotal object profiles
after deblending, such that each ACS pixel can only be assigned to one
object (or the background).  For objects which are not detected in
ACS, but which are detected in IRAC, we simply use the original IRAC
SExtractor photometry.  Because of the larger IRAC beam, for all
objects in the catalog we impose a minimum semi-major axis of
2\arcsec.  In all cases aperture corrections are computed individually
from PSFs provided by the SSC based on the aperture sizes and shapes
used for
photometry\footnote{http://ssc.spitzer.caltech.edu/irac/psf.html}.

\begin{figure}
\plotone{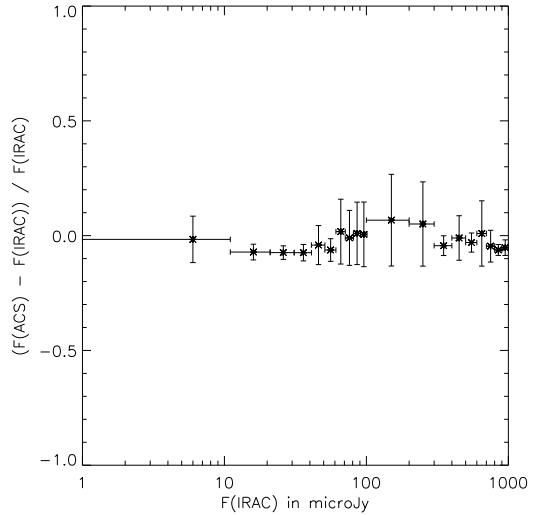}
\caption[photerr]{Difference between ACS based (F(ACS)) and IRAC-alone
  based (F(IRAC)) photometry for the 18,300 sources in common in the
  catalog with flux measurements in greater than five wavebands.
  Points and error bars show the biweight center and dispersion of the
  distrubution within each bin.  This allows for robust outlier
  rejection.  Note that bin size changes with flux.  }
\label{fig:photerr}
\epsscale{1}
\end{figure}

Final aperture photometry was performed using custom extraction
software written in IDL and based on the APER and MASK$\_$ELLIPSE
routines with the shape information from SExtractor, from either ACS
or IRAC as described above, using local backgrounds.  Fluxes are
measured in all IRAC bands at the location of all ACS detections
without applying a flux limit (IRAC detection limits are discussed
below).  Because we use local backgrounds, the measured fluxes of
objects near the confusion limit should have a larger scatter than
those non-confused objects, but will on average be the correct flux.
This will not effect the photometric redshifts, as it will likely
shift all IRAC points up or down, but not relative to each other.
Figure \ref{fig:photerr} shows the difference between using the IRAC
photometry alone and using the ACS based photometry as described here.
This plot is a direct analog of Figure 2 in \citet{laidler2007} which
shows the comparison of fluxes for the software package TFIT.  We do
not see the systematic offset in fluxes that they see at low flux
levels.



\begin{figure}
\plotone{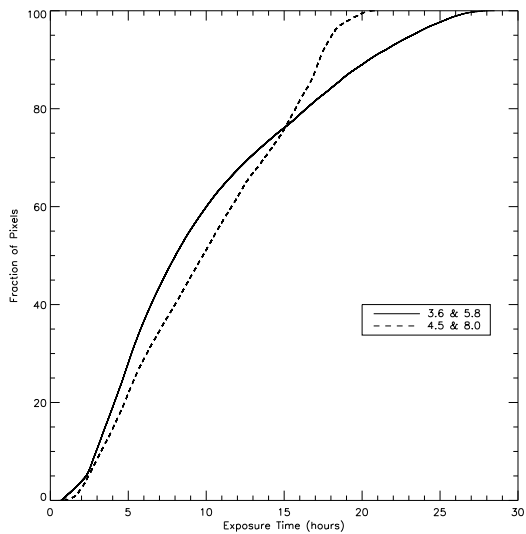}
\caption[ch1cov]{Cumulative distribution of the number of pixels
  versus exposure time in hours for the central $15\arcmin$ diameter
  region.  The solid line shows the coverage for the 3.6 \& 5.8 \micron\
  images, the dashed line is for 4.5 \& 8.0 \micron\ images. }
\label{fig:cov}
\epsscale{1}
\end{figure}

Determining the detection limits of the IRAC data is complicated by
varying exposure times across the field, source confusion, and our use
of ACS locations as priors for photometry.  Field coverage is a
complex function of spacecraft dynamics, see Figure \ref{fig:cov} and
\ref{fig:ch1}.  Within the central 180 square arcminutes of the field,
the median exposure time per pixel is eight hours at 3.6 \& 5.8
microns and ten hours at 4.5 \& 8.0 microns.  The maximum exposure
time in the center is 28 hours and 21 hours at 3.6 \& 5.8 and 4.5 \&
8.0 microns respectively.  A total of 40 hours of observations went
into covering the entire field.  The confusion limit in the IRAC data
arises because both the PSF (2\arcsec {\sc FWHM}) and the number of
sources is large.  We have used ACS priors to be able to deblend
confused sources, and therefore our detection limit is below the
nominal confusion limit.  One effect of using ACS for deblending is
that the depth of our photometry will be color dependent.  We have
IRAC photometry below the confusion limit for all objects which are
detected in the F814W band, but cannot deblend objects for which their
is no F814W counterpart.  Therefore, our photometry is deeper for
sources which have bluer spectral shapes, whereas it is confusion
limited for the redder, IRAC only detections.  Because of these three
complexities, there is no one single value for the detection limit of
the survey.

\begin{figure}
\plotone{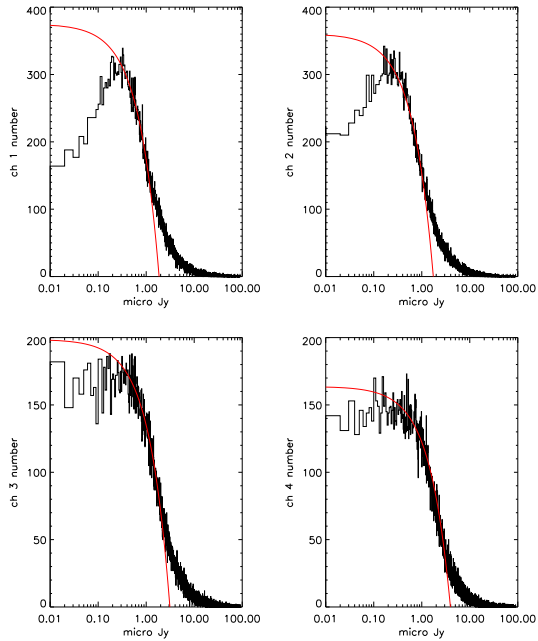}
\caption[iraccomplete]{Number counts in the IRAC bands showing the completeness
  limits for each band.  The histogram is the measured number of
  sources and the solid (red) line is a linear fit to the rising portion of the
  histogram.  Completeness is determined where the measured number of
  sources drops below 95\% of the expected number of sources based on
  the fits.}
\label{fig:iraccomplete}
\epsscale{1}
\end{figure}

We are able to measure nominal $95\%$ completeness limits in the IRAC
passbands from a number count diagram by measuring where the number of
observed objects drops below 95\% of the expected number of objects.
The expected numbers are calculated by fitting a straight line to the
brighter flux number counts and assuming that the faint number counts
would also follow that line in the absence of confusion.  Figure
\ref{fig:iraccomplete} shows the number counts in the four IRAC bands
with a best fit line to the detections to show where the number of
measured fluxes drop below 95\% of the expected numbers.  The figure
also shows the larger confusion in the shorter wavelength bands.  The
measured and expected number counts are lower in the longer wavelength
bands.  Final values at 3.6, 4.5, 5.8, and 8.0\,\micron\ are 0.2,
0.17, 0.11, 0.11\,$\mu$Jy respectively (see Table \ref{tab:data}).  A
more complete analysis of the number counts will be presented in a
future paper.


\subsubsection{MIPS 70\micron}
The 70\micron\ observations were taken with the Multiband Imaging
Photometer for {\it Spitzer} \citep[MIPS;][]{rieke2004} with a total
exposure time per pixel of 1.6 hours covering 160 square arcminutes to
a limiting $4\sigma$ depth of 4.3 mJy (Program ID 30222).  The data were taken
as a 5$\times$2 grid of pointings ($5.3\arcmin\times2.6\arcmin$ field of
view) in small scale photometry mode with 10 second exposures. The raw
70\micron\ data were downloaded from the SSC archive and were reduced
using the Germanium Reprocessing Tools (GeRT, version 20060415).  The
pipeline was optimized using the same parameters derived for the
ultra-deep GOODS-North photometry data \citep{frayer2006b}.  We used a
combination of the column median filter and a high-pass median time
filter per detector to remove the instrumental artifacts
("streaking").  The filtering was done in two passes to remove the
filtering artifacts ("negative side-lobes") around the brightest
sources \citep{frayer2006a}.  A calibration factor of 702
MJy\,sr$^{-1}$ per MIPS-70-unit \citep{gordon2007} was assumed, and
the data have been color corrected by dividing by 0.918
\citep{stansberry2007}.  This color-correction factor is for a
constant $\nu\,S_{\nu}$ SED across the filter bandpass, but is
appropriate to within 2\% for a wide range of galaxy SEDs ($S_{\nu}
\propto \nu^{-\alpha}$, $\alpha =0$--3).

The data were combined using the SSC mosaicing software ({\sc
  mopex}).  PSF fitting has been found to be more accurate for MIPS
70\micron\ data than aperture photometry \citep{gordon2007}.  Source
extraction was done using the Astronomical Point-Source Extraction
({\sc apex}) tools within {\sc mopex}. The 70 \micron\ peaks were
fitted using the Point-source Response Function (PRF; FWHM =
18\farcs6) image derived from the extragalactic First Look Survey
\citep{frayer2006a} because there are not enough isolated bright
sources in the field to accurately derive our own PRF.  Sixty-one
sources were extracted with a signal-to-noise ratio (SNR) greater than
four (4.3mJy).  To avoid spurious sources at 70 \micron\, only
those 42 sources with 24 \micron\ counterparts were included in this
study.  We don't expect any sources to be detected at 70\micron\ that
are not also detected at 24\micron\ which we confirm by eye.  The
remaining unmatched sources are blends that we are unable to separate
or noise peaks.  We applied an aperture correction of 1.15 for
emission outside of the adopted PRF \citep{frayer2009}.

\subsubsection{MIPS 24\micron}

The 24\micron\ data were taken with MIPS on {\it Spitzer} with a total
exposure time of one hour per pixel covering 225 square arcminutes to
a limiting $3\sigma$ depth of 17.3 $\mu$ Jy (Program ID 1868).  The data were taken in
large-field photometry mode with a 30-second exposure time. A
$3\times3$ MIPS field of view grid was mapped and repeated five times,
with multiple dithers and chops. There were a total of 1080 separate
exposures with a final total depth of 60 minutes per pointing on the
sky.  The MIPS data were processed by the {\it Spitzer} Science Center
into individual image BCDs (pipeline version S12.4). However, substantial ``jailbar''
artifacts, as well as a significant gradient, were visible. All of the
individual pointings were forced to a common background by applying an additive
constant to the entire frame. A ``delta-dark'' was then generated from
the median of all frames; the great degree of dithering in the data
allows this process to reject all actual celestial objects in the
frames from the median stack. That stack was then adjusted to a median
overall zero value, and then subtracted from all the data. It
currently is not known whether the gradient effect is additive or
multiplicative, although our experience with other Si:As arrays of
this kind strongly suggests (from a physical basis) that it is
additive. However, we reduced the data both ways, and found no
difference at any detectable level. The data were then coadded using
the {\sc mopex} software package onto exactly the same projection
system as used for IRAC, albeit with $1\farcs2$ pixels.

IRAF {\sc daofind} was used for object detection.  We supply the code
with the PSF FWHM (5\farcs8) and background sigma values taken by
examining the image. {\sc Daofind} then counts the flux within an
annulus of diameter FWHM and flags any set of pixels as a detection
where that flux is above a threshold of five sigma.  To deal with
confused sources, we perform object detection iteratively.  After the
first run through {\sc daofind}, all objects are subtracted from the
image using a PSF determined from the detected objects. {\sc Daofind}
is then re-run on the residual image.  Due to the large PSF and the
lack of nearby galaxies in the field, with the exception of only a
handful of galaxies, all MIPS detections appear as point sources.
Photometry on all detected sources is done with the IRAF task {\sc
  allstar} which fits PSF's to groups of objects simultaneously.  An
aperture correction of 1.4 is applied for flux beyond the 6.5 pixel
radius at which the PSF star was normalized.  This correction factor
is calculated from a curve of growth based on the composite PSF star.
Using this method the $3\sigma$ detection limit is 17.3$\mu$Jy.  These
noise properties are comparable to the GOODS slightly longer exposure
(77 minute) dataset that has a $3\sigma$ limit of 12$\mu$Jy.

\subsection {Akari Space Telescope}
The 4, 11, 15, and 18\micron\ data were taken with the Infrared Camera
\citep[IRC;][]{onaka2007} on the {\it Akari} Space Telescope
\citep{murakami2007} with a total exposure time of 1 - 2.5 hours per
band per pixel covering 100 square arcminutes to a $3\sigma$ depth of 1.5, 15,
33, and 45 $\mu$Jy respectively.  The details of this data set will be
provided in Egami et al. (2009, in preparation), so here we only
describe some basic data parameters.

The IRC was used to obtain images at 11, 15, and 18 \micron\ during
2006, October.  These three Akari/IRC bands bridge the {\em Spitzer}
imaging wavelength gap between IRAC 8 $\mu$m and MIPS 24 \micron.  A 4
\micron\ image was also obtained since the 4 and 11 \micron\ data were
taken simultaneously for the same region by design.  Each image covers
a field of view of $\sim$10\arcmin$\times$10\arcmin\ (i.e., four times
larger than that of IRAC) with a pixel scale of 2\farcs3--2\farcs5.
The field center coordinates are 17:40:00 $+$69:00:00 (J2000).

Low-resolution infrared spectra were also taken for sources in the
same 10\arcmin$\times$10\arcmin\ area using the slit-less multi-object
spectroscopy (MOS) mode of IRC \citep{ohyama2007}.  When combined,
obtained spectra cover a wavelength range of 1.8--26.5
\micron\footnote{There is a gap in coverage at 13.4-17.5 \micron\ because
the grism covering this range has been damaged.} with a resolving
power of 20--50.  The integration times were several hours.


\subsection{Palomar and KPNO near-IR}

The J, H, and K$_s$ observations were taken with the WIRC camera
\citep{wilson2003} on the Palomar 200-inch Hale telescope and the
FLAMINGOS camera on the KPNO 4m telescope. Total exposure times at
Palomar were 0.8, 0.7, and 1.6 hours per pixel covering 120, 120, and
210 square arcminutes to a 3$\sigma$ limiting depth of 21.8, 22.1, and
21.2 mags (AB) in J, H, and K$_s$ respectively.  KPNO was used solely
for the J band with a total exposure time of 0.5 hours per pixel
covering 220 square arcminutes to a 3$\sigma$ limiting depth of 21.4
mag (AB).  Palomar data were taken on 2006 May 5, 2008 May 15-18, and
2008 June 27-28 and the KPNO data were taken on 2006 August 29-30.  In
both cases the large format cameras are well-matched to the field size
allowing efficient observing.

The Palomar and KPNO data were reduced using a custom IDL pipeline
(available from the authors).  Blank sky images are generated from the
actual survey data to serve as a temporally windowed measure of the
dark current and sky.  An initial pass through the data generates a
first-order image reduction on which SExtractor is used to detect and
mask objects and bad pixels within the data. A second pass utilizes
the generated masks to reject pixels from the stacking process that
creates the blank sky images. These ``skydarks'' were then subtracted
from the data. Additional tasks repair other image defects, such as
quadrant-level floating bias uncertainties. The data were flat-fielded
using dome flats. Each frame was individually calibrated both
photometrically and astrometrically to 2MASS, which compensates for
drifts in calibration due to non-photometric conditions. Individual
images were re-projected using
SWARP\footnote{$http://terapix.iap.fr/rubrique.php?id\_rubrique=49$}
onto a common projection center used for all the other datasets. The
data were then coadded using IRAF/IMCOMBINE, with relative weighting
of the images based on their noise properties. The photometric and
astrometric calibration of the final coadded mosaic was then further
refined by a final rematch to the 2MASS survey. An object catalog was
extracted using SExtractor, with a detection threshold of
2$\sigma$. The SExtractor MAG\_AUTO photometric magnitude was
used. Seeing was typically 1.1-1.5\arcsec.  In general, the WIRC data
are used for the survey as that dataset is deeper than the FLAMINGOS dataset. WIRC
$3\sigma$ detection limits vary from center (22.5,23.1,22.5) to the
edge (21.8,22.1,21.2) in J, H, K (AB) respectively. The FLAMINGOS
3$\sigma$ detection limit is 21.4 mag (AB) in the J--band.

\subsection {MMT near-IR}
The {\sl J}- and {\sl H}-band observations were taken with the SAO
Widefield InfraRed Camera \citep[SWIRC;][]{brown2008} at the 6.5m MMT
with a total exposure time of 7.1 and 2.1 hours covering 25 and 225
square arcminutes to a 3$\sigma$ limiting depth of 24.4 and 23.0
respectively.  The data were taken over the course of two observing
runs. The first set of observations were taken during the four nights
of 2005 April 19--22.  Conditions were not photometric, and the field
was observed through thin cirrus.  Total integration times were 14
minutes at both {\sl J}- and {\sl H}-band at each of nine positions
arranged in an overlapping $3\times3$ configuration of a
5.1\arcmin$\times$5.1\arcmin field of view to cover roughly the central
$15\arcmin\times15\arcmin$ of the IRAC dark field.  The total integration
time was built up out of short, dithered 60\,s ({\sl J}) and 30\,s
({\sl H}) exposures.  The seeing was variable but remained below
1\arcsec\ FWHM for most of the run.  The second set of observations
were carried out during the three nights of 2006 May 12--14.  During
this campaign only the central $5\arcmin\times5\arcmin$ of the IRAC cal
field (the subregion with the deepest overall IRAC integration), was
observed, and only at {\sl J}-band.  A total of 257 70\,s dithered
exposures were acquired through variable cirrus with sub-arcsecond
seeing for a total combined exposure time of 5 hours.

The individual exposures were processed using standard methods.  All
frames were dark- and bias-subtracted using dark frames taken at the
beginnings and ends of the corresponding nights.  A custom mask was
constructed using a median stack of all the 60\,s {\sl J}-band
exposures from the 2005 run, so as to exclude large regions of bad and
noisy pixels from subsequent analysis.  The celestial coordinates,
which were initially only accurate to within 20\arcsec, were refined
and verified with the {\tt wcstools} package by reference to bright
2MASS stars.  The images were
flattened with skyflats.  This was an iterative process whereby a
crude, initial mosaic was first constructed to reveal faint objects
not easily detected in the individual exposures.  Objects in this
initial mosaic were then identified and masked, and the deeper object
masks were subsequently translated to the coordinate frames of all
individual frames.  In this way, with the much deeper object masks,
much-improved skyflats were created from the stack of individual
frames.  Specifically, this technique eliminated the negative
artifacts that arose from repeated placement of faint-but-significant
objects on the same parts of the detector array by the dither pattern.

Because none of the data were taken on photometric nights, the
photometric calibrations were referenced to an ensemble of
well-detected 2MASS stars.  Final mosaics were constructed using
0\farcs15 pixels by coadding the individual frames using sigma
clipping to eliminate cosmic rays.  Two wide, shallow
$15^\prime\times15$\arcmin\ mosaics were made having 3$\sigma$ depths
of 22.7 ({\sl J}) and 23.0 ({\sl H}) AB magnitudes, based on the noise
measured within a 3\arcsec\ diameter aperture.  In the smaller but
deeper mosaic from 2006 a 3$\sigma$ depth of 24.4 AB mag was estimated
within the same aperture.  The FWHM in the final mosaics was roughly
0\farcs75.

Photometry was measured in the final mosaics with SExtractor, using
MAG\_AUTO as a measure of total magnitude, thus not requiring an
aperture correction.  For the wide-field mosaics, SExtractor was used
in two-image mode.  In this case, the IRAC 3.6\,\micron\ mosaic was
used as the detection image, and the {\sl J}- and {\sl H}-band mosaics
were used as the measurement images.  For the deep {\sl J}-band
mosaic, SExtractor was used in single-image mode.  Given the 0\farcs15
pixels and the 0\farcs75 seeing, six connected pixels each
individually detected at the 1.5\,$\sigma$ level were required for a
source detection.  The resulting deep {\sl J}-band catalog contains
over 13,000 sources.


\subsection {MMT optical}

The z$^\prime$ observations were taken with the Megacam camera at the
6.5m MMT with a total exposure time of 3.6 hours per pixel covering
580 square arcminutes to a 3$\sigma$ limiting depth of 25.9 mag (AB).
The observations consisted of a series of 500\,s exposures taken on
the nights of 2007 June 1-4 under photometric conditions, except for
the night of 2007 June 1, when some thin cirrus was present. Seeing
varied from as low as 0\farcs6 to as high as 1\farcs7, and averaged
about 1\farcs0.  The data were reduced interactively using standard
techniques on a dedicated pipeline machine hosted at SAO.
Bias-subtracted and flattened exposures were treated to remove cosmic
rays and bad pixels before calculating the coordinates using stars
from the USNOB1.0 catalog and correcting the photometry for off-axis
scattered light.  The resulting exposures were spatially registered to
a common coordinate system.  All frames taken on the night of 2007
June 1 were then flux-calibrated using exposures from the photometric
nights.  All the frames were then scaled appropriately and coadded to
create the final z$^\prime$ mosaic. The mosaic covers roughly
$24\arcmin\times24\arcmin$ covering the entire IRAC survey field where
the center has the deepest exposure time of 3.6 hours tapering towards
the field edge.

The $z^\prime$ mosaic was photometered with SExtractor in single-image
mode, using MAG-AUTO as a total magnitude.  Because the seeing was
measured to be approximately 1\farcs0 in the combined image, the
detection algorithm was configured to identify all objects having at
least 12 connected 0\farcs32 pixels lying 1$\sigma$ above the
background.  Using the coverage map, portions of the mosaic observed
three or fewer times were masked, so that photometry was measured only
for sources observed at least four times by Megacam.  We estimate that
the depth reached is 25.9 AB mag, 3$\sigma$, within a 2\arcsec
aperture.


\subsection{HST / ACS Optical}

{\it HST} / ACS F814W observations were taken with a total exposure
time of 1.4 hours per pixel covering 260 square arcminutes to a
3$\sigma$ limiting depth of 28.6 mags (AB) (Program ID 10521).  ACS
WFC is a mosaic of two $2048\times4096$ SITE CCDs with a pixel scale
of 0.05\arcsec/pixel ($15\mu$ pixels), so that the full field of view
per exposure is 11.3 square arcminutes. Observations were made over 50
orbits between November 28, 2006 and December 07, 2006 using SIDE-2
electronics.  The observations were divided into 25 pointings, and
each pointing was in turn divided into eight roughly 10 minute
exposures for a total of 200 exposures.  We used a small dither
between exposures at each pointing for cosmic ray rejection and to
cover the gap between the two ACS CCDs (2.5\arcsec).  We also employ a
$\sim 0.5\arcmin$ overlap between pointings.  The entire ACS mosaic is
contained within the {\it Spitzer} IRAC map.

The ACS pipeline {\it calacs} was used for basic reduction of the
images including overscan and bias subtraction, cosmic ray rejection,
dark subtraction, and flat fielding.  Beyond this base reduction we
developed our own method of further reducing and mosaicing the 25
pointings. The three largest challenges in doing this were the large
amount of memory required to combine 200 160Mb fits files, getting the
astrometry correct given distortions and relative shifts between
images caused by thermal effects in the telescope, and the variation
in background level due to the four amplifiers used to readout the two
CCDs.  {\sc multidrizzle} alone is unable to account for these
effects.  Our pipeline uses a combination of the IRAF tasks {\sc
  multidrizzle}, {\sc tweakshifts}, and {\sc ccmap} along with
automatic background subtraction by quadrant and {\sc SWarp} for the
final mosaicing.  The final combined mosaic image is 1.7GB in size and
covers a $\sim 20\arcmin$ diameter rough circle in the center of the
IRAC dark field with the native 0\farcs05 per pixel resolution and 0\farcs1 seeing.

The photometry zeropoint was taken from the headers.  SExtractor was
used with the same basic setup as the Palomar optical data for source
detection and extraction (see \S \ref{pal_optical}).  The SExtractor
MAG\_AUTO is used in the catalog which uses a Kron total magnitude
when there are no near neighbors to bias the measurement and a
corrected isophotal magnitude otherwise.  The $3\sigma$ detection
limit for point sources is $I = 28.6(AB)$.  There are $\sim 51,000$
sources in the catalog with detections in both ACS and IRAC.


\subsection{Palomar Optical}
\label{pal_optical}

The u$^\prime$, g$^\prime$, r$^\prime$, and i$^\prime$ data were taken
with the Large Format Camera at the 200'' Hale telescope at Palomar
with a total exposure time of 6.0, 6.2, 4.1, and 4.7 hours per pixel covering
450 square arcminutes to a 3$\sigma$ limiting depth of 27.2, 27.1,
26.6, and 26.0 mags (AB) respectively.  Observing runs occurred on
2004 July 22-24 and 2004 August 08-10.  All observing nights had less
than 50\% moon illumination.  The LFC is a mosaic of six
$2048\times4096$ CCDs with a pixel scale of 0.18\arcsec/pixel ($15\mu$
pixels), so that the full field of view per exposure is roughly a
$24\arcmin $ diameter circle.  Images were randomly dithered between
exposures to account for the gap between chips (15\arcsec).
Integration times were typically 360 seconds in r$^\prime$ and
i$^\prime$, 900 seconds in u$^\prime$, and 600 seconds in g$^\prime$.
u$^\prime$-band science and calibration data were binned $2\times2$
pixels.

Reduction of this Palomar data follows standard procedures.  The
images were cross-talk and overscan corrected using IRAF tasks.  A
combined bias frame was then made from roughly five overscan corrected
individual bias frames and applied to the data to remove the large
scale bias.  No dark current correction was required.
Pixel--to--pixel sensitivity variations were corrected in all images
using 15 median-combined dome flats taken throughout the run with
7000-9000 counts each.  Although twilight flats were taken during the
August run, they were not required in the reduction.  i'-band images
were fringe corrected using a median smoothed, object removed,
bad-pixel masked combination of all science frames.

The IRAF tasks {\sc mscgetcatalog}, {\sc msccmatch} and {\sc
  msccsetwcs} were used to find and apply x and y shifts and rotations
to each frame by tying it to the USNO-A2 positions.  Fits were done
manually by ruthlessly rejecting outliers while maintaining even
coverage over the field.  A combined bad pixel mask was made for all
cosmic rays(CR), hot pixels, and satellite trails for each image using
{\sc craverage} and masking by hand.  The bad pixel mask was used with
the IRAF task {\sc fixpix} which does a linear interpolation over the
effected pixels.  To remove gradients in the background we use {\sc
  mscskysub} to fit a second order Legendre to the background of each
image and subtract it.  SWARP was then used to re-project and combine
together all images in each filter.

Photometric calibration was performed in the usual manner using a
field of SDSS well measured stars at a range of airmasses.
Photometric nights were analyzed together; solutions were found in
each filter for a common magnitude zero-point.  All nights except
August 10 were photometric.  Those exposures taken in non-photometric
conditions were individually tied to the photometric data using
roughly 10 stars well distributed around each frame.

We use SExtractor to both find all objects in the combined frames, and
to determine their shape parameters.  The detection threshold was
defined such that objects have a minimum of six contiguous pixels,
each of which are greater than $1.5\sigma$ above the background sky
level.  We choose these parameters as a compromise between detecting
faint objects in high signal-to-noise regions and rejecting noise
fluctuations in low signal-to-noise regions.  This corresponds to
3$\sigma$ point source detection limits of 27.2, 27.1, 26.6, and 26.0 mags
(AB) in u$^\prime$, g$^\prime$, r$^\prime$, and i$^\prime$ respectively.
Shape parameters are determined in SExtractor using only those pixels
above the detection threshold.

When making the catalog, we use an aperture magnitude for objects
fainter than 22.5 and a Kron magnitude from SExtractor for all objects
brighter than 22.5.  We choose a kron factor of 2.5 and a minimum
radius of 3.0 pixels as can be shown to give the closest estimate to
total magnitudes \citep{bertin1996}.  The fainter objects are all
point sources and so do not benefit from shape information in their
photometry.  Aperture sizes are 3\farcs6, 5\farcs6, 4\farcs1, and
4\farcs5 for u$^\prime$, g$^\prime$, r$^\prime$, and i$^\prime$
respectively.  Aperture sizes were chosen based on a curve of growth
of a bright star in the field.  We choose large apertures ($\sim
4\times$ FWHM) to enclose 'all' of the flux and therefore not require
an aperture correction.


\subsection{Palomar Optical Spectroscopy}

Spectroscopy was done with the COSMIC spectrograph at the Palomar Hale
200'' telescope.  COSMIC, at prime focus, has a 13.6\arcmin\ field of
view, and 0.4\arcsec\ pixels.  Observations were made over four nights
in June of 2007 \& 2008 with the 300 l/mm grating with a dispersion of
3 \AA\ per pixel.  We chose a slit-width of 1.5\arcsec\ to match our 1
- 1.5\arcsec\ seeing.  The optical band covered by this instrument,
roughly 3500 to 9000\AA, includes such spectral features as CaH\&K,
[OII], [OIII], H$\alpha$, H$\beta$, H$\delta$, G band, and the 4000
\AA break. During both runs we were able to observe a total of 11
slitmasks of $\sim 25$ galaxies each with exposure times of on average
80 minutes divided into multiple exposures.  One HgAr lamp and one
flat exposure was taken through each mask at the beginning of the
night for calibration.  Targets were selected to be galaxies
(non-stellar PSF's) brighter than r=21(AB) with priority in mask
design given to those with MIPS 24 or 70\micron\ detections to boost
the chance of seeing an emission line and thereby getting a secure
redshift.

Reduction was done with IRAF mainly through the {\sc Bogus2006}
\footnote{https://zwolfkinder.jpl.nasa.gov\slash ~stern\slash
  homepage\slash bogus.html} scripts.  Prior to running {\sc bogus},
images were overscan and bias subtracted.  {\sc Bogus} itself does a
2D reduction including flat-fielding, cosmic ray removal, sky
subtraction, fringe suppression and combination of frames.  The same
reduction is performed on both science images and arcs.  The standard
IRAF tasks of {\sc apall, identify}, and {\sc dispcor} were used to
wavelength correct, trace, and extract the spectra with a secondary
background subtraction for minor level changes.  One dimensional
spectra were extracted for a total of 200 galaxies with measurable
continuum. Although spectroscopic standards were observed during each
run, we did not flux calibrate the spectra as we are only interested
in determining redshifts (see \S \ref{photz}).

\subsection{Chandra ACIS-I}

{\it Chandra} ACIS-I observations were taken with a total exposure
time of 28 hours per pixel covering 290 square arcminutes to a
limiting depth of $3.6\times 10^{-16}$ergs/s/cm$^2$.  The camera is a
$2\times2$ mosaic of four $1024\times1024$ frontside--illuminated CCDs
arranged to be tangent to the mirror assembly in the focal plane.  The
pixel scale is 0.49\arcsec/pixel ($24\mu$ pixels), so that the full
field of view per exposure is roughly 290 square arcminutes.
Observing was done in three separate observations at different
pointing angles to reduce the effect of the gap between chips
(11\arcsec); 50ks on 2007 July 23, 22ks on 2007 September 28, and 28ks
on 2007 September 29, for a total exposure time of 100ks (observation
ID's \dataset[ADS/Sa.CXO#Obs/7359]{7359},
\dataset[ADS/Sa.CXO#Obs/8471]{8471},
\dataset[ADS/Sa.CXO#Obs/9595]{9595}).  Observations were done in timed
exposure mode ({\sc TE}) with very faint format ({\sc VFAINT}) in
order to measure events in $5\times5$ pixel islands instead of smaller
regions which has the advantage of better determination of background
events such as cosmic rays.  We know from the ROSAT survey that there
are no bright objects in the field such that pile-up of flux would be
a problem in these larger event regions.  The data were pre-processed
by the {\it Chandra} X-ray center with standard data processing (SDP)
version 7.6.11.1.  Further reduction was done with the {\it Chandra}
Interactive Analysis of Observations software (CIAO 4.0).  This
version of the SDP includes identifying and correcting for hot pixels
and cosmic ray afterglows ( residual charge on frames after a cosmic
ray hit).

In creating an event list we include background cleaning with the code
{\sc acis\_process\_events} which flags possible background events
based on counts in the outer ring of the $5\times 5$ event island so
that they can be removed in the filtering phase.  We then filter the
event list for the number of pixels above a threshold and flags for
bad pixels or cosmic rays.  Secondly, frames which are not considered
good science frames due either to the telescope motion, angle, or
temperature are removed from the event list. Lastly we filter the data
on energy range, keeping only energies between 0.3 and 7 KeV where the
particle background is fairly flat and quantum efficiency is the
highest.

Because the observations were taken in three different exposures we
correct the astrometry before combining individual observations into
an exposure map and fluxed image.  To do this we use a preliminary run
of {\sc wavdetect} to make a source list and then {\sc
  reproject\_aspect} to reproject the images to a reference image (in
this case the first, longest observation ID 7359).  {\sc merge\_all}
is used to create the final combined image and exposure map.  After
combination we examine the data for background flares and remove
anomalous high background times from the image.  We perform source
detection and photometry on the full 0.3 - 7 Kev band data using {\sc
  wavdetect} which operates by correlating the data with ``mexican
hat'' wavelet functions on size scales from one to eight pixels.  The
significance threshold parameter in {\sc wavdetect} is set such that
we expect approximately one false detection per $1000\times1000$ pixel
area, or approximately four in our field.  There are 121 total source
detections.

Photometry with {\sc wavdetect} sums the total counts within the
source cell defined by the shape which encompasses all positive count
pixels around the source detection.  Backgrounds are determined in the
wavelet analysis.  To convert from count rate to flux units we assume
an average spectral shape of $\Gamma =1.4$ \citep{giacconi2002,
  elvis2009,laird2009}.  Along with the local galactic NH value of
$4\times10^{20}cm^{-2}$ this gives us energy conversion factors of
$5.49\times10^{-12}$, $1.8\times10^{-11}$, and $1.08\times10^{-11}$ in
the soft, hard, and total bands respectively. Using $\Gamma =1.7$
\citep{manners2003} changes these conversion factors by $\sim 10\%$.
The minimum number of counts we have detected in the full band is
$3.3\pm 2.0$, corresponding to $3.6\pm2.2\times10^{-16}$erg/cm$^2$/s.

\section{The Catalog}
\label{catalog}
The catalog is built by matching sources from each band to
the remainder of the catalog within a radius that is similar to that
band's PSF size.  The disadvantages of this method are miss-matches due
to noise detections and deblending imperfections.  A noise detection
specifically in the ACS band will propagate through to the IRAC bands
because of our photometry methods (see Section \ref{irac}).  Saturated
stars with diffraction spikes often produce spurious signals.
Fortunately this field has few/no bright stars by design (see Section
\ref{intro}). The brightest star in the field is 12.2 mag (AB) is
F814W.  The main source of error in the matching will be incorrect
matches because of multiple sources within the match radius.  For this
reason we match catalogs with similar PSF sizes.  For those bands with
especially large PSFs we are careful to match sources with other large
PSF bands, and not directly to our smallest PSF catalog.  Specifically
the MIPS 24\micron\ dataset is matched to the IRAC source list, as the
catalog nearest in PSF size to itself.  Secondly the MIPS 70\micron\
dataset is matched to the MIPS 24\micron\ data.  This method means
that we will lose any MIPS 24\micron\ sources which are not detected
at IRAC wavelengths and similarly any 70\micron\ sources which are not
detected at 24\micron.  We expect these objects to be extremely rare,
if they exist at all, and therefore they will be searched for in a
different manner.

For this catalog we do not use matched apertures, but instead
we calculate a ``total'' flux in each band.  We do not want to lose
information by smoothing all the bands to the same resolution, but
instead keep all information possible at each wavelength.  To this
end, we use aperture corrections where necessary, as discussed above.

By far, the {\it HST} ACS F814W catalog has the most number of sources
and so dominates the spatial distribution of sources.  As mentioned in
\S \ref{irac}, the ACS catalog is used as the detection image for the
IRAC data.  Other optical and near-IR bands are matched to that merged
catalog of ACS and IRAC.  In total there are 86,815 sources in the
matched catalog.  This catalog will be made publicly available at a
future date.  At this time all of the space-based data are available
from their respective archives.

\begin{figure}
\plotone{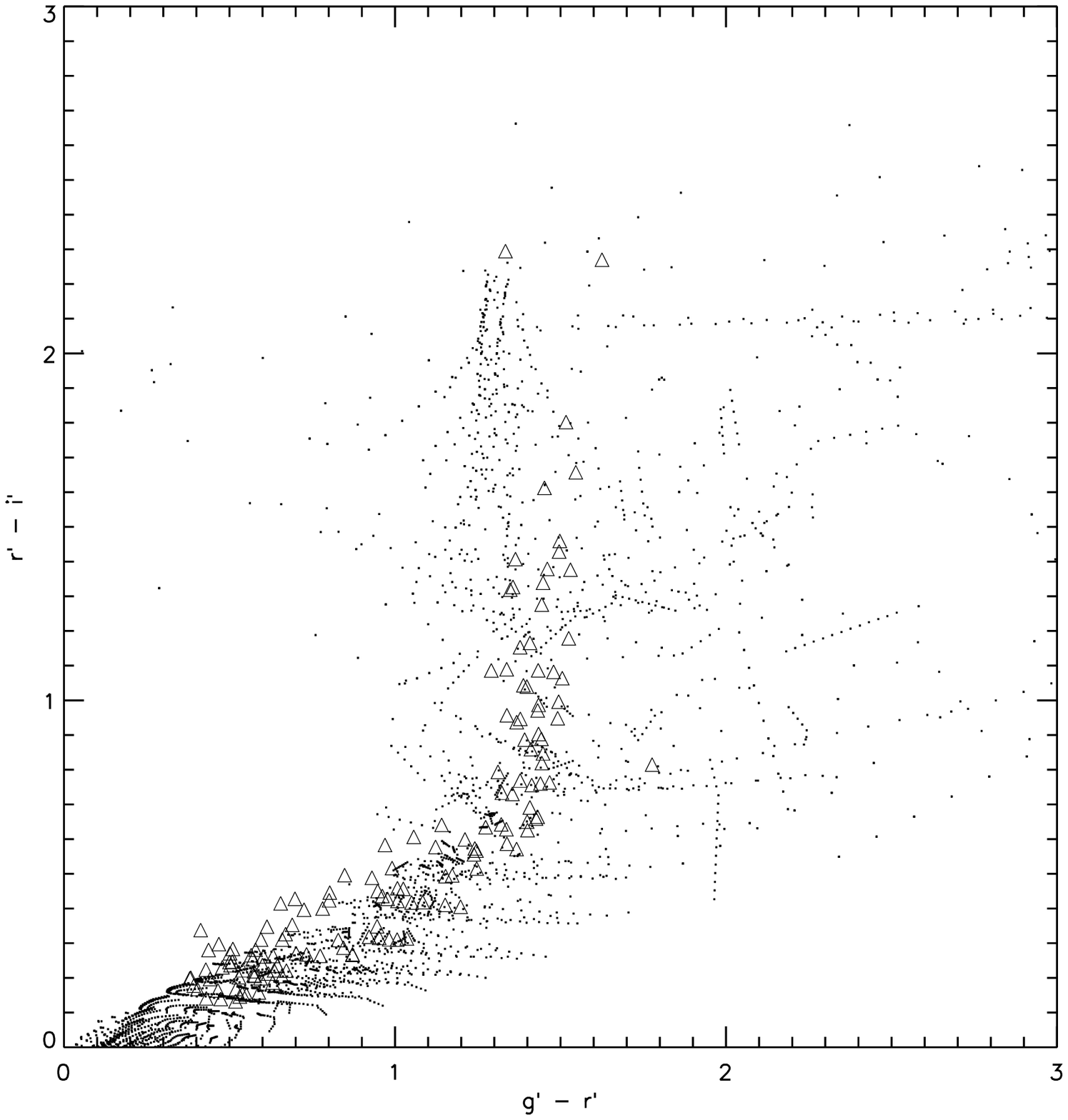}
\plotone{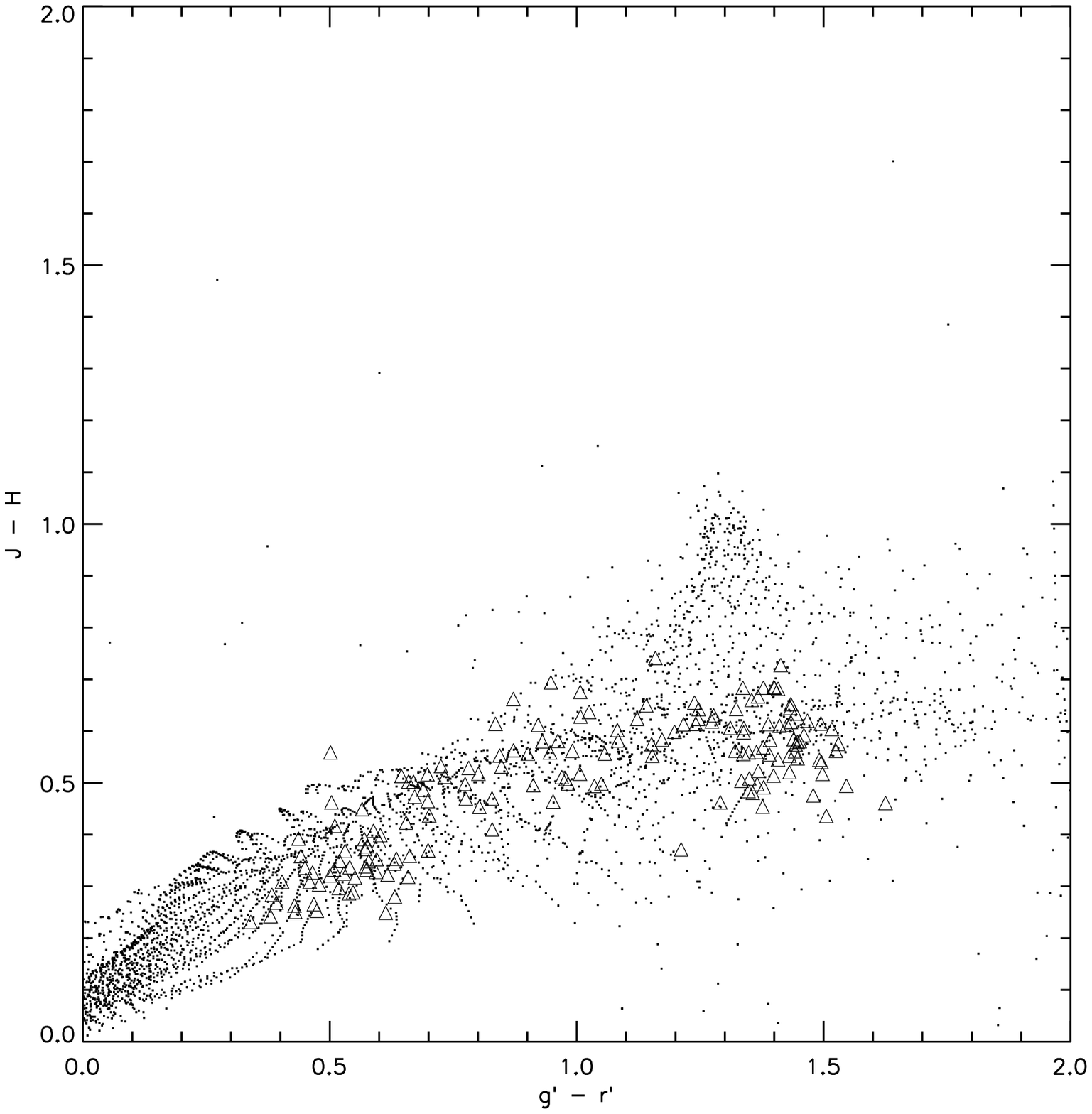}
\caption[stars]{Optical and near-IR color color diagrams as a check of
  photometry.  Colors from synthetic stellar spectra from
  \citet{lejeune2001} for a set of stars covering a range in mass and
  metallicity are shown as black points.  The 150 stars from the dark
  field are shown as triangles.  }
\label{fig:stars}
\epsscale{1}
\end{figure}

As a check of the catalog photometry, we present color-color diagrams
of stars in the field.  Stars are a good population on which to test
our photometry since they follow well defined tracks in color space.
We select stars from our ACS catalog.  A plot of the SExtractor {\sc
  isoarea} parameter as a function of magnitude clearly separates a
ridge of unresolved point sources from the resolved sources because
unresolved sources have smaller {\sc isoarea} than galaxies at any
given magnitude.  Although we have used our highest resolution data to
pick out the stars, there are undoubtedly interlopers from higher
redshift compact galaxies or quasars.  In total we have $\sim 150$
point sources with detections in u$^\prime$, g$^\prime$, r$^\prime$,
i$^\prime$, J, and H.  In Figure \ref{fig:stars} we compare the optical
and near-IR colors of these stars to those from synthetic stellar
spectra from the BaSeL library \citep{lejeune2001}.  The comparison
sample includes synthetic stars with masses from 0.4-0.8 to 120-150
$M\odot$ and metallicities Z=0.0005 to 0.1 measured at ages between 1000
yr and 16-20 Gyr.  We find good agreement between the real star colors
and those from the theoretical models.

\section{Photometric Redshifts}
\label{photz}

The combined IRAC and ACS catalog contains over $50,000$ objects which
makes acquisition of spectroscopic redshifts impractical.  Even
confirmation spectroscopy of red galaxies at $z=1$ in our three
candidate clusters will require many nights on 8-10m class telescopes
and is therefore also impractical.  In lieu of spectroscopy we use our
extensive multi-wavelength, broad-band catalog to build spectral
energy distributions (SEDs) using up to 13 bands (u', g', r', i', F814W,
z$^\prime$, J, H, K, 3.6, 4.5, 5.8, 8.0\micron) from which we derive
photometric redshifts.

\begin{figure}
\plotone{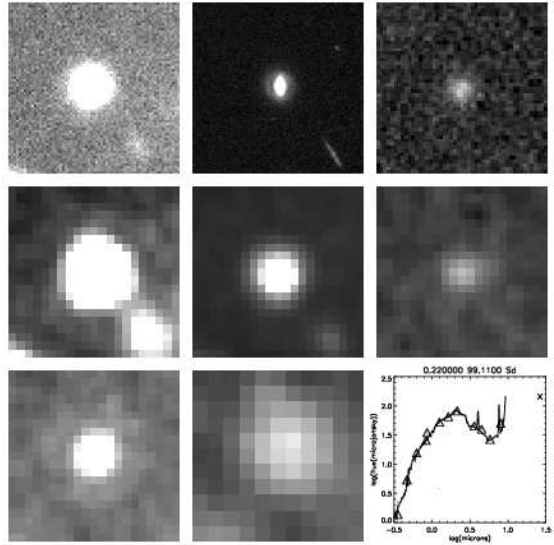}
\caption[photz]{Example spiral galaxy with a well fit photometric
  redshift by hyperz.  The thumbnails show a $12\arcsec\times 12\arcsec$ box in
  $r^{\prime}$, HST / ACS F814W, K, 3.6, 4.5, 5.8, 8.0, 24 \micron, and
  the fitted SED.  The title of the SED plot gives the best fit
  photometric redshift, the probability that that redshift is good,
  and the name of the best-fit template.  Triangles are data used in
  the fit, the line is the fit, and the cross is the 24 \micron\
  data point, not used in the hyperz fit, but plotted here for
  completeness.}
\label{fig:photz}
\epsscale{1}
\end{figure}

These SEDs are fit with template spectra from \citet{Polletta2007}
\footnote{$http://www.iasf-milano.inaf.it/~polletta/templates/swire\_templates.html$}.
These templates have been used successfully by a number of surveys at
a range of redshifts for all galaxy types
\citep{adami2008,negrello2009,salvato2009,ilbert2009}.  Because the
templates include the IR spectral region (empirically built from
observations in SWIRE), we find them the best choice to use as models
for this dataset.  We use 15 templates including ellipticals, spirals,
star forming galaxies, and AGN.  Photometric redshifts are calculated
using {\sc Hyperz}; a chi-squared minimization fitting program
including a correction for interstellar reddening
\citep{bolzonella2000, calzetti2000}. An example of a well fit galaxy
is shown in Figure \ref{fig:photz}. Errors in photometric redshifts
are determined by comparing the photometric redshifts with
spectroscopic redshifts.

Spectroscopic redshifts were determined using both IRAF tasks {\sc
  emsao} and {\sc xcsao}.  Specifically {\sc emsao} searches the
spectrum for both absorption and emission lines which it correlates
with a given line list.  {\sc xcsao} cross-correlates the spectrum
with known galaxy templates which allows us to use features like the
4000 \AA\ break and the rest of the spectral shape to identify
redshifts.  Both techniques were used together to arrive at the best
fit redshift for each galaxy. We used 17 spectral templates of
galaxies and AGN from the compilation of the {\it HST} Calibration
Database System \citep{francis1991, kinney1996, calzetti1994}.  We do
not use the SWIRE SED templates here (as we do for the photometric
redshifts above) because the optical spectra in the SWIRE templates is
based on modeling and their true strength lies in their IR treatment.
Since our spectroscopy is in the optical, we use the Calibration
Database System templates because they are based more reliably on UV
and optical observations.  We applied a very strict requirement that
all emission and absorption features in the 1D spectra were confirmed
by eye on the 2D spectra and that multiple lines be identified in all
cases to avoid incorrect redshift determination due to cosmic rays or
noise features from sky line subtraction.

\begin{figure}
\plotone{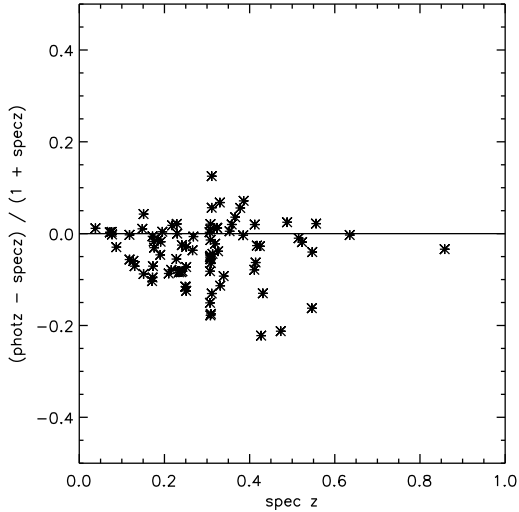}
\caption[specvsphotz]{Comparison of spectroscopically and
  photometrically determined redshifts for 87 galaxies.  The scatter
  implies an error on the photometric redshifts of $0.064(1+z)$.}
\label{fig:specvsphotz}
\epsscale{1}
\end{figure}

We were able to successfully determine redshifts for 87 galaxies.
This represents a conservative sample of 'good' redshift
determinations defined to have either high signal-to-noise emission
lines or multiple absorption lines and good cross correlations.  We
then compare the spectroscopic to photometric redshifts to quantify
the error on the latter (Figure \ref{fig:specvsphotz}). There are
cases where {\sc Hyperz} has failed to fit the correct redshift.  In
these galaxies it appears that {\sc Hyperz} has not chosen a template
that goes through any of the photometric points, completely missing
the SED shape.  This could be due to a failure of hyperz or the
errors in the photometry.  Those galaxies, as characterized by a
$\chi^{2}$ value greater than 50, are not included in this comparison
 sample.  The error on the photometric redshifts
is $0.064(1+z)$.  Note that this error is quoted as a function of
redshift and so takes into account the increasing scatter with z.
This accuracy is similar to other IRAC based multi-wavelength studies
\citep{brodwin2006, rowanrobinson2008}.

\section{Initial Science Results}
\label{discuss}

Given this unique dataset, we are carrying out scientific
programs to exploit the available information.  \citet{Krick2008} have
discovered a large scale structure at redshift one consisting of three
galaxy clusters.  These clusters show evidence of a build-up of the
red cluster sequence at redshift one.  In a study of the star
formation and AGN fraction in those clusters, \citet{krick2009} have
found evidence for star formation in star forming galaxies to look the
same regardless of environment, but a dense environment does
strengthen the suppression of star formation and AGN activity in
clusters over the field from redshift one to the present.

Ongoing work continues on finding and characterizing high redshift
galaxies ($z > 6$), compton thick AGN, the lowest mass brown dwarfs,
population III SNe \citep{frost2009}, the reddest objects in the
universe from their $24\micron - r$ colors, and other interesting
phenomenon, the study of which is uniquely suited to this field.  As
this is the only mid-IR dataset with time domain information on a 5+
year time-line, we are exploring the variable and transient objects in
the field, from stars to AGN \citep[][and Foesneau et al. 2009, in
preparation]{hund2006, surace2007}.  One such application is to study
the extragalactic variability in mid--IR to set constraints on the
nature of dust around AGN.

\begin{figure}
\plotone{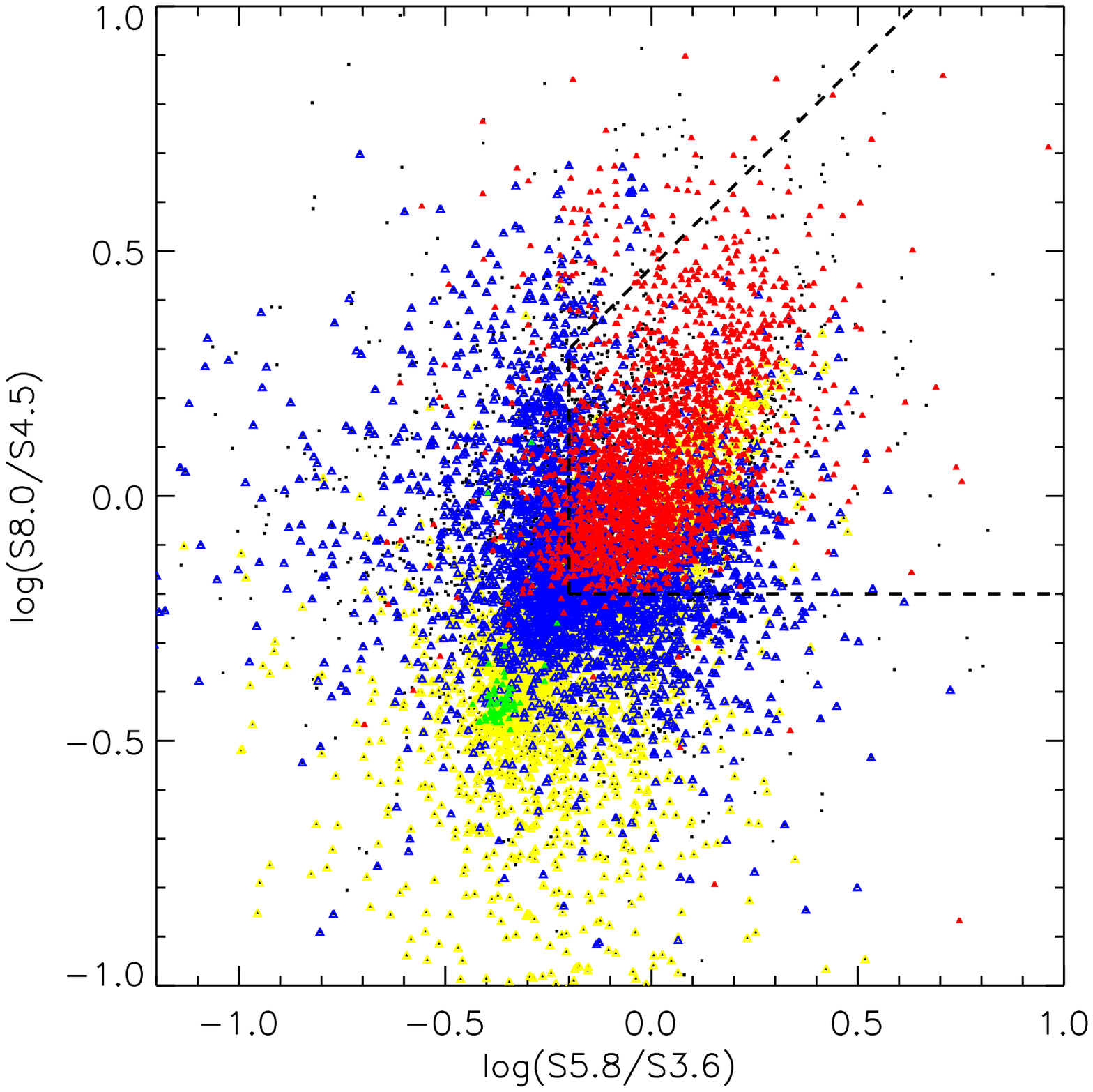}
\plotone{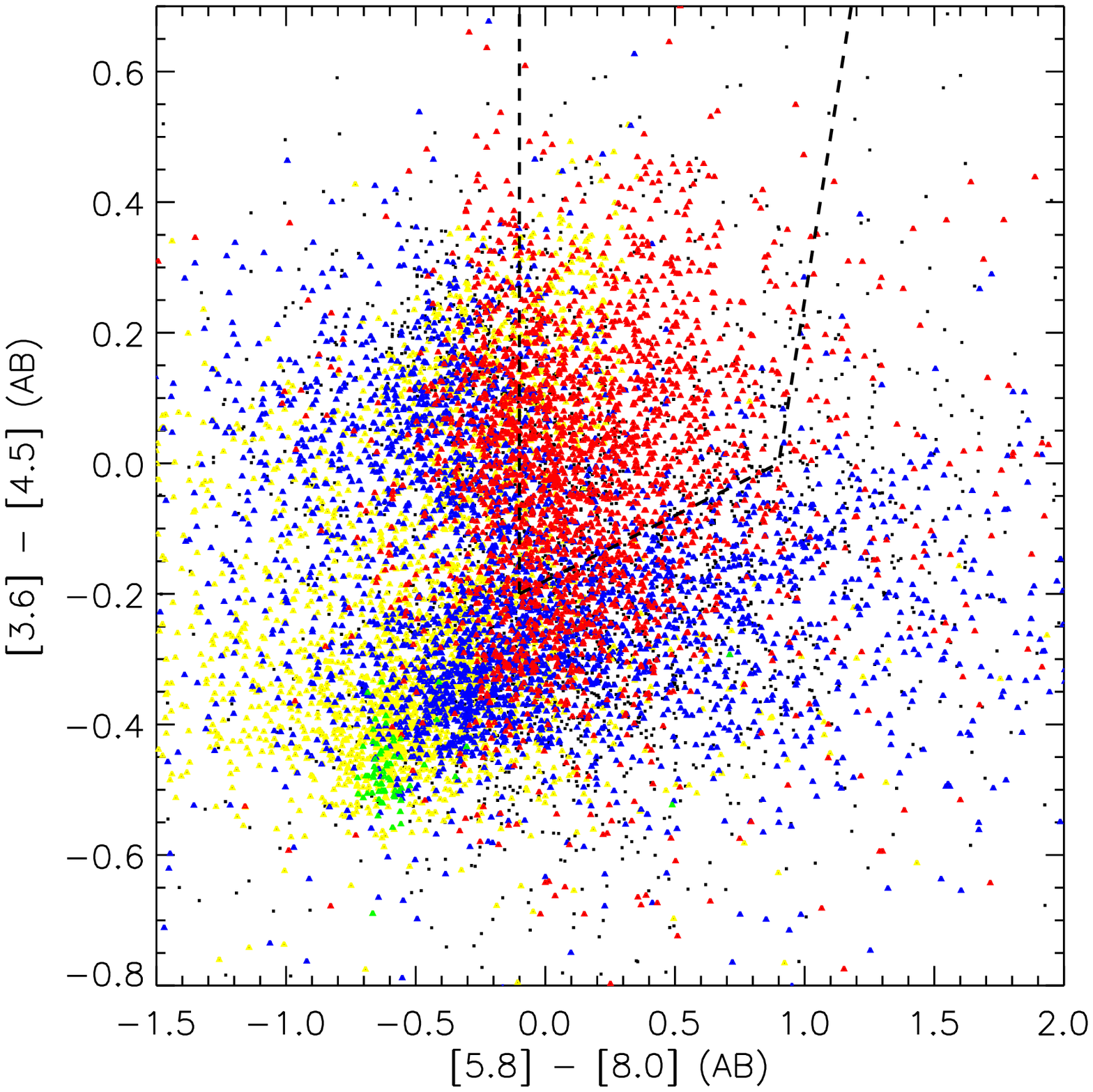}
\caption[lacycolor]{IRAC color color diagram in the fashion of a)
  \citet{lacy2004}, b) \citet{stern2005}.  Sources are color coded by
  best fit SED shape.  Green, yellow, blue, and red indicate stars,
  ellipticals, spirals, and AGN.  While the \citet{lacy2004} wedge has
  high completeness for the AGN sample, there is significant
  contamination, and the \citet{stern2005} wedge shows lower
  completeness but less contamination.}
\label{fig:iraccolorcolor}
\epsscale{1}
\end{figure}

As a representation of the data, we include here a discussion of IRAC
color color diagrams to both differentiate between AGN and star
forming galaxies \citep{lacy2004, stern2005} and demonstrate the
effectiveness of our ACS based IRAC photometry.  Fig
\ref{fig:iraccolorcolor} shows all sources originally detected in the
four IRAC bands before adding ACS shape information.  Here we have
separated the sample based on SED classifications into a spiral,
elliptical, AGN, and stellar sample.  Those SEDs best fit with
composite templates are not colored in the plot.  Deep surveys like
the IRAC dark field, in opposition to the shallower, wide-area,
surveys like the {\it Spitzer} first look survey and the IRAC shallow
survey \citep{eisenhardt2004} show a larger amount of contamination in
the AGN wedges from intermediate redshift spiral and elliptical
galaxies \citep{sajina2005, barmby2006, barger2008}.  Those AGN which
are blue-ward (left) of the \citet{lacy2004} wedge are AGN at redshift
$\sim 1.0$ that have a stellar peak in addition to a power law shape,
and that stellar peak has been shifted into the IRAC 3.6 \micron\
band.  We see from these plots that our SED fitting produces nice
results for the morphologies of the galaxies.

Secondly we discuss the effect of our ACS based IRAC photometry on the
scatter in the color color diagrams.  We do see larger scatter than
the original \citet{lacy2004} and \citet{stern2005} diagrams because
this is a deeper survey.  The larger scatter is also seen in other
deep surveys like that in the Extended Groth Strip \citep{barmby2006}.
We use this scatter in the color color diagrams to quantify the
difference between using the IRAC only photometry and the ACS based
IRAC photometry.  We use the same sample of $\sim 11,000$ galaxies
originally detected in all four IRAC bands to make color-color
distributions of both IRAC only and ACS based IRAC photometry.  The
change in scatter is quantified with the F-test where the ratio of
variances gives probabilities that the scatters in the distributions
are the same.  Those probabilities are less than $1\times10^{-14}$
indicating that the scatter is statistically significantly reduced if
we use the ACS based photometry instead of the IRAC only photometry.

\acknowledgments
 We thank the anonymous referee for useful suggestions on the
 manuscript.  This research has made use of data from the Two Micron All Sky Survey,
which is a joint project of the University of Massachusetts and the
Infrared Processing and Analysis Center/California Institute of
Technology, funded by the National Aeronautics and Space
Administration and the National Science Foundation.  This work was
based on observations obtained with the Hale Telescope, Palomar
Observatory as part of a continuing collaboration between the
California Institute of Technology, NASA/JPL, and Cornell University,
the {\it Spitzer} Space Telescope, which is operated by the Jet
Propulsion Laboratory, California Institute of Technology under a
contract with NASA, and the NASA/ESA Hubble Space Telescope, obtained
at the Space Telescope Science Institute, which is operated by the
Association of Universities for Research in Astronomy, Inc.  Support
for program \#10521 was provided by NASA through a grant from the
Space Telescope Science Institute, which is operated by the
Association of Universities for Research in Astronomy, Inc., under
NASA contract NAS 5-26555.  FLAMINGOS was designed and constructed by
the IR instrumentation group (PI: R. Elston) at the University of
Florida, Department of Astronomy, with support from NSF grant
AST97-31180 and Kitt Peak National Observatory.  Observations reported
here were obtained at the MMT Observatory, a joint facility of the
Smithsonian Institution and the University of Arizona.  Support for
this work was provided by the National Aeronautics and Space
Administration through Chandra Award Number G07-8120 issued by the
Chandra X-ray Observatory Center, which is operated by the Smithsonian
Astrophysical Observatory for and on behalf of the National
Aeronautics Space Administration under contract NAS8-03060.

{\it Facilities:} \facility{Hale (LFC, WIRC, COSMIC)}, \facility{MMT(Megacam, SWIRC)}, \facility{HST (ACS)}, \facility{Spitzer (IRAC, MIPS)},
\facility{Akari}, \facility{CXO (ACIS)}, \facility{KPNO()}


\bibliography{ms.astroph.bbl}  

\begin{thebibliography}{55}
\expandafter\ifx\csname natexlab\endcsname\relax\def\natexlab#1{#1}\fi

\bibitem[{{Adami} {et~al.}(2008){Adami}, {Ilbert}, {Pell{\'o}}, {Cuillandre},
  {Durret}, {Mazure}, {Picat}, \& {Ulmer}}]{adami2008}
{Adami}, C., {Ilbert}, O., {Pell{\'o}}, R., {Cuillandre}, J.~C., {Durret}, F.,
  {Mazure}, A., {Picat}, J.~P., \& {Ulmer}, M.~P. 2008, \aap, 491, 681

\bibitem[{{Barger} {et~al.}(2008){Barger}, {Cowie}, \& {Wang}}]{barger2008}
{Barger}, A.~J., {Cowie}, L.~L., \& {Wang}, W.-H. 2008, \apj, 689, 687

\bibitem[{{Barmby} {et~al.}(2006){Barmby}, {Alonso-Herrero}, {Donley}, {Egami},
  {Fazio}, {Georgakakis}, {Huang}, {Laird}, {Miyazaki}, {Nandra}, {Park},
  {P{\'e}rez-Gonz{\'a}lez}, {Rieke}, {Rigby}, \& {Willner}}]{barmby2006}
{Barmby}, P., {et~al.} 2006, \apj, 642, 126

\bibitem[{{Bertin} \& {Arnouts}(1996)}]{bertin1996}
{Bertin}, E., \& {Arnouts}, S. 1996, \aaps, 117, 393

\bibitem[{{Bolzonella} {et~al.}(2000){Bolzonella}, {Miralles}, \&
  {Pell{\'o}}}]{bolzonella2000}
{Bolzonella}, M., {Miralles}, J.-M., \& {Pell{\'o}}, R. 2000, \aap, 363, 476

\bibitem[{{Brodwin} {et~al.}(2006){Brodwin}, {Brown}, {Ashby}, {Bian}, {Brand},
  {Dey}, {Eisenhardt}, {Eisenstein}, {Gonzalez}, {Huang}, {Jannuzi},
  {Kochanek}, {McKenzie}, {Murray}, {Pahre}, {Smith}, {Soifer}, {Stanford},
  {Stern}, \& {Elston}}]{brodwin2006}
{Brodwin}, M., {et~al.} 2006, \apj, 651, 791

\bibitem[{{Brown} {et~al.}(2008){Brown}, {McLeod}, {Geary}, \&
  {Bowsher}}]{brown2008}
{Brown}, W.~R., {McLeod}, B.~A., {Geary}, J.~C., \& {Bowsher}, E.~C. 2008, in
  Presented at the Society of Photo-Optical Instrumentation Engineers (SPIE)
  Conference, Vol. 7014, Society of Photo-Optical Instrumentation Engineers
  (SPIE) Conference Series

\bibitem[{{Calzetti} {et~al.}(2000){Calzetti}, {Armus}, {Bohlin}, {Kinney},
  {Koornneef}, \& {Storchi-Bergmann}}]{calzetti2000}
{Calzetti}, D., {Armus}, L., {Bohlin}, R.~C., {Kinney}, A.~L., {Koornneef}, J.,
  \& {Storchi-Bergmann}, T. 2000, \apj, 533, 682

\bibitem[{{Calzetti} {et~al.}(1994){Calzetti}, {Kinney}, \&
  {Storchi-Bergmann}}]{calzetti1994}
{Calzetti}, D., {Kinney}, A.~L., \& {Storchi-Bergmann}, T. 1994, \apj, 429, 582

\bibitem[{{Eisenhardt} {et~al.}(2004){Eisenhardt}, {Stern}, {Brodwin}, {Fazio},
  {Rieke}, {Rieke}, {Werner}, {Wright}, {Allen}, {Arendt}, {Ashby}, {Barmby},
  {Forrest}, {Hora}, {Huang}, {Huchra}, {Pahre}, {Pipher}, {Reach}, {Smith},
  {Stauffer}, {Wang}, {Willner}, {Brown}, {Dey}, {Jannuzi}, \&
  {Tiede}}]{eisenhardt2004}
{Eisenhardt}, P.~R., {et~al.} 2004, \apjs, 154, 48

\bibitem[{{Elvis} {et~al.}(2009){Elvis}, {Civano}, {Vignali}, {Puccetti},
  {Fiore}, {Cappelluti}, {Aldcroft}, {Fruscione}, {Zamorani}, {Comastri},
  {Brusa}, {Gilli}, {Miyaji}, {Damiani}, {Koekemoer}, {Finoguenov}, {Brunner},
  {Urry}, {Silverman}, {Mainieri}, {Hasinger}, {Griffiths}, {Carollo}, {Hao},
  {Guzzo}, {Blain}, {Calzetti}, {Carilli}, {Capak}, {Ettori}, {Fabbiano},
  {Impey}, {Lilly}, {Mobasher}, {Rich}, {Salvato}, {Sanders}, {Schinnerer},
  {Scoville}, {Shopbell}, {Taylor}, {Taniguchi}, \& {Volonteri}}]{elvis2009}
{Elvis}, M., {et~al.} 2009, ArXiv e-prints

\bibitem[{{Fazio} {et~al.}(2004){Fazio}, {Hora}, {Allen}, {Ashby}, {Barmby},
  {Deutsch}, {Huang}, {Kleiner}, {Marengo}, {Megeath}, {Melnick}, {Pahre},
  {Patten}, {Polizotti}, {Smith}, {Taylor}, {Wang}, {Willner}, {Hoffmann},
  {Pipher}, {Forrest}, {McMurty}, {McCreight}, {McKelvey}, {McMurray}, {Koch},
  {Moseley}, {Arendt}, {Mentzell}, {Marx}, {Losch}, {Mayman}, {Eichhorn},
  {Krebs}, {Jhabvala}, {Gezari}, {Fixsen}, {Flores}, {Shakoorzadeh}, {Jungo},
  {Hakun}, {Workman}, {Karpati}, {Kichak}, {Whitley}, {Mann}, {Tollestrup},
  {Eisenhardt}, {Stern}, {Gorjian}, {Bhattacharya}, {Carey}, {Nelson},
  {Glaccum}, {Lacy}, {Lowrance}, {Laine}, {Reach}, {Stauffer}, {Surace},
  {Wilson}, {Wright}, {Hoffman}, {Domingo}, \& {Cohen}}]{fazio2004}
{Fazio}, G.~G., {et~al.} 2004, \apjs, 154, 10

\bibitem[{{Fern{\'a}ndez-Soto} {et~al.}(1999){Fern{\'a}ndez-Soto}, {Lanzetta},
  \& {Yahil}}]{fernandez-soto1999}
{Fern{\'a}ndez-Soto}, A., {Lanzetta}, K.~M., \& {Yahil}, A. 1999, \apj, 513, 34

\bibitem[{{Francis} {et~al.}(1991){Francis}, {Hewett}, {Foltz}, {Chaffee},
  {Weymann}, \& {Morris}}]{francis1991}
{Francis}, P.~J., {Hewett}, P.~C., {Foltz}, C.~B., {Chaffee}, F.~H., {Weymann},
  R.~J., \& {Morris}, S.~L. 1991, \apj, 373, 465

\bibitem[{{Frayer} {et~al.}(2006{\natexlab{a}}){Frayer}, {Fadda}, {Yan},
  {Marleau}, {Choi}, {Helou}, {Soifer}, {Appleton}, {Armus}, {Beck}, {Dole},
  {Engelbracht}, {Fang}, {Gordon}, {Heinrichsen}, {Henderson}, {Hesselroth},
  {Im}, {Kelly}, {Lacy}, {Laine}, {Latter}, {Mahoney}, {Makovoz}, {Masci},
  {Morrison}, {Moshir}, {Noriega-Crespo}, {Padgett}, {Pesenson}, {Shupe},
  {Squires}, {Storrie-Lombardi}, {Surace}, {Teplitz}, \&
  {Wilson}}]{frayer2006a}
{Frayer}, D.~T., {et~al.} 2006{\natexlab{a}}, \aj, 131, 250

\bibitem[{{Frayer} {et~al.}(2006{\natexlab{b}}){Frayer}, {Huynh}, {Chary},
  {Dickinson}, {Elbaz}, {Fadda}, {Surace}, {Teplitz}, {Yan}, \&
  {Mobasher}}]{frayer2006b}
---. 2006{\natexlab{b}}, \apjl, 647, L9

\bibitem[{{Frayer} {et~al.}(2009){Frayer}, {Sanders}, {Surace}, {Aussel},
  {Salvato}, {Le Floc'h}, {Huynh}, {Scoville}, {Afonso-Luis}, {Bhattacharya},
  {Capak}, {Fadda}, {Fu}, {Helou}, {Ilbert}, {Kartaltepe}, {Koekemoer}, {Lee},
  {Murphy}, {Sargent}, {Schinnerer}, {Sheth}, {Shopbell}, {Shupe}, \&
  {Yan}}]{frayer2009}
---. 2009, ArXiv e-prints

\bibitem[{{Frost} {et~al.}(2009){Frost}, {Surace}, {Moustakas}, \&
  {Krick}}]{frost2009}
{Frost}, M.~J., {Surace}, J.~A., {Moustakas}, L.~A., \& {Krick}, J.~E. 2009,
  \apjl, 698, L68

\bibitem[{{Giacconi} {et~al.}(2001){Giacconi}, {Rosati}, {Tozzi}, {Nonino},
  {Hasinger}, {Norman}, {Bergeron}, {Borgani}, {Gilli}, {Gilmozzi}, \&
  {Zheng}}]{giacconi2001}
{Giacconi}, R., {et~al.} 2001, \apj, 551, 624

\bibitem[{{Giacconi} {et~al.}(2002){Giacconi}, {Zirm}, {Wang}, {Rosati},
  {Nonino}, {Tozzi}, {Gilli}, {Mainieri}, {Hasinger}, {Kewley}, {Bergeron},
  {Borgani}, {Gilmozzi}, {Grogin}, {Koekemoer}, {Schreier}, {Zheng}, \&
  {Norman}}]{giacconi2002}
---. 2002, \apjs, 139, 369

\bibitem[{{Giavalisco} {et~al.}(2004){Giavalisco}, {Ferguson}, {Koekemoer},
  {Dickinson}, {Alexander}, {Bauer}, {Bergeron}, {Biagetti}, {Brandt},
  {Casertano}, {Cesarsky}, {Chatzichristou}, {Conselice}, {Cristiani}, {Da
  Costa}, {Dahlen}, {de Mello}, {Eisenhardt}, {Erben}, {Fall}, {Fassnacht},
  {Fosbury}, {Fruchter}, {Gardner}, {Grogin}, {Hook}, {Hornschemeier}, {Idzi},
  {Jogee}, {Kretchmer}, {Laidler}, {Lee}, {Livio}, {Lucas}, {Madau},
  {Mobasher}, {Moustakas}, {Nonino}, {Padovani}, {Papovich}, {Park},
  {Ravindranath}, {Renzini}, {Richardson}, {Riess}, {Rosati}, {Schirmer},
  {Schreier}, {Somerville}, {Spinrad}, {Stern}, {Stiavelli}, {Strolger},
  {Urry}, {Vandame}, {Williams}, \& {Wolf}}]{giavalisco2004}
{Giavalisco}, M., {et~al.} 2004, \apjl, 600, L93

\bibitem[{{Gordon} {et~al.}(2007){Gordon}, {Engelbracht}, {Fadda},
  {Stansberry}, {Wachter}, {Frayer}, {Rieke}, {Noriega-Crespo}, {Latter},
  {Young}, {Neugebauer}, {Balog}, {Beeman}, {Dole}, {Egami}, {Haller}, {Hines},
  {Kelly}, {Marleau}, {Misselt}, {Morrison}, {P{\'e}rez-Gonz{\'a}lez}, {Rho},
  \& {Wheaton}}]{gordon2007}
{Gordon}, K.~D., {et~al.} 2007, \pasp, 119, 1019

\bibitem[{{Grazian} {et~al.}(2006){Grazian}, {Fontana}, {de Santis}, {Nonino},
  {Salimbeni}, {Giallongo}, {Cristiani}, {Gallozzi}, \&
  {Vanzella}}]{grazian2006}
{Grazian}, A., {et~al.} 2006, \aap, 449, 951

\bibitem[{{Hund} {et~al.}(2006){Hund}, {Ashby}, {Hora}, \& {Surace}}]{hund2006}
{Hund}, L.~B., {Ashby}, M.~L., {Hora}, J.~L., \& {Surace}, J. 2006, in Bulletin
  of the American Astronomical Society, Vol.~38, Bulletin of the American
  Astronomical Society, 1095--+

\bibitem[{{Ilbert} {et~al.}(2009){Ilbert}, {Capak}, {Salvato}, {Aussel},
  {McCracken}, {Sanders}, {Scoville}, {Kartaltepe}, {Arnouts}, {LeFloc'h},
  {Mobasher}, {Taniguchi}, {Lamareille}, {Leauthaud}, {Sasaki}, {Thompson},
  {Zamojski}, {Zamorani}, {Bardelli}, {Bolzonella}, {Bongiorno}, {Brusa},
  {Caputi}, {Carollo}, {Contini}, {Cook}, {Coppa}, {Cucciati}, {de la Torre},
  {de Ravel}, {Franzetti}, {Garilli}, {Hasinger}, {Iovino}, {Kampczyk},
  {Kneib}, {Knobel}, {Kovac}, {LeBorgne}, {LeBrun}, {LeF{\`e}vre}, {Lilly},
  {Looper}, {Maier}, {Mainieri}, {Mellier}, {Mignoli}, {Murayama}, {Pell{\`o}},
  {Peng}, {P{\'e}rez-Montero}, {Renzini}, {Ricciardelli}, {Schiminovich},
  {Scodeggio}, {Shioya}, {Silverman}, {Surace}, {Tanaka}, {Tasca}, {Tresse},
  {Vergani}, \& {Zucca}}]{ilbert2009}
{Ilbert}, O., {et~al.} 2009, \apj, 690, 1236

\bibitem[{{Kinney} {et~al.}(1996){Kinney}, {Calzetti}, {Bohlin}, {McQuade},
  {Storchi-Bergmann}, \& {Schmitt}}]{kinney1996}
{Kinney}, A.~L., {Calzetti}, D., {Bohlin}, R.~C., {McQuade}, K.,
  {Storchi-Bergmann}, T., \& {Schmitt}, H.~R. 1996, \apj, 467, 38

\bibitem[{{Krick} {et~al.}(2008){Krick}, {Surace}, {Thompson}, {Ashby}, {Hora},
  {Gorjian}, \& {Yan}}]{Krick2008}
{Krick}, J.~E., {Surace}, J.~A., {Thompson}, D., {Ashby}, M.~L.~N., {Hora},
  J.~L., {Gorjian}, V., \& {Yan}, L. 2008, \apj, 686, 918

\bibitem[{{Krick} {et~al.}(2009){Krick}, {Surace}, {Thompson}, {Ashby}, {Hora},
  {Gorjian}, \& {Yan}}]{krick2009}
---. 2009, ArXiv e-prints

\bibitem[{{Lacy} {et~al.}(2004){Lacy}, {Storrie-Lombardi}, {Sajina},
  {Appleton}, {Armus}, {Chapman}, {Choi}, {Fadda}, {Fang}, {Frayer},
  {Heinrichsen}, {Helou}, {Im}, {Marleau}, {Masci}, {Shupe}, {Soifer},
  {Surace}, {Teplitz}, {Wilson}, \& {Yan}}]{lacy2004}
{Lacy}, M., {et~al.} 2004, \apjs, 154, 166

\bibitem[{{Laidler} {et~al.}(2007){Laidler}, {Papovich}, {Grogin}, {Idzi},
  {Dickinson}, {Ferguson}, {Hilbert}, {Clubb}, \& {Ravindranath}}]{laidler2007}
{Laidler}, V.~G., {et~al.} 2007, \pasp, 119, 1325

\bibitem[{{Laird} {et~al.}(2009){Laird}, {Nandra}, {Georgakakis}, {Aird},
  {Barmby}, {Conselice}, {Coil}, {Davis}, {Faber}, {Fazio}, {Guhathakurta},
  {Koo}, {Sarajedini}, \& {Willmer}}]{laird2009}
{Laird}, E.~S., {et~al.} 2009, \apjs, 180, 102

\bibitem[{{Lejeune} \& {Schaerer}(2001)}]{lejeune2001}
{Lejeune}, T., \& {Schaerer}, D. 2001, \aap, 366, 538

\bibitem[{{Lonsdale} {et~al.}(2003){Lonsdale}, {Smith}, {Rowan-Robinson},
  {Surace}, {Shupe}, {Xu}, {Oliver}, {Padgett}, {Fang}, {Conrow},
  {Franceschini}, {Gautier}, {Griffin}, {Hacking}, {Masci}, {Morrison},
  {O'Linger}, {Owen}, {P{\'e}rez-Fournon}, {Pierre}, {Puetter}, {Stacey},
  {Castro}, {Polletta}, {Farrah}, {Jarrett}, {Frayer}, {Siana}, {Babbedge},
  {Dye}, {Fox}, {Gonzalez-Solares}, {Salaman}, {Berta}, {Condon}, {Dole}, \&
  {Serjeant}}]{lonsdale2003}
{Lonsdale}, C.~J., {et~al.} 2003, \pasp, 115, 897

\bibitem[{{Makovoz} \& {Marleau}(2005)}]{makovoz2005}
{Makovoz}, D., \& {Marleau}, F.~R. 2005, \pasp, 117, 1113

\bibitem[{{Manners} {et~al.}(2003){Manners}, {Johnson}, {Almaini}, {Willott},
  {Gonzalez-Solares}, {Lawrence}, {Mann}, {Perez-Fournon}, {Dunlop}, {McMahon},
  {Oliver}, {Rowan-Robinson}, \& {Serjeant}}]{manners2003}
{Manners}, J.~C., {et~al.} 2003, \mnras, 343, 293

\bibitem[{{Murakami} {et~al.}(2007){Murakami}, {Baba}, {Barthel}, {Clements},
  {Cohen}, {Doi}, {Enya}, {Figueredo}, {Fujishiro}, {Fujiwara}, {Fujiwara},
  {Garcia-Lario}, {Goto}, {Hasegawa}, {Hibi}, {Hirao}, {Hiromoto}, {Hong},
  {Imai}, {Ishigaki}, {Ishiguro}, {Ishihara}, {Ita}, {Jeong}, {Jeong},
  {Kaneda}, {Kataza}, {Kawada}, {Kawai}, {Kawamura}, {Kessler}, {Kester},
  {Kii}, {Kim}, {Kim}, {Kobayashi}, {Koo}, {Kwon}, {Lee}, {Lorente}, {Makiuti},
  {Matsuhara}, {Matsumoto}, {Matsuo}, {Matsuura}, {M{\"u}ller}, {Murakami},
  {Nagata}, {Nakagawa}, {Naoi}, {Narita}, {Noda}, {Oh}, {Ohnishi}, {Ohyama},
  {Okada}, {Okuda}, {Oliver}, {Onaka}, {Ootsubo}, {Oyabu}, {Pak}, {Park},
  {Pearson}, {Rowan-Robinson}, {Saito}, {Sakon}, {Salama}, {Sato}, {Savage},
  {Serjeant}, {Shibai}, {Shirahata}, {Sohn}, {Suzuki}, {Takagi}, {Takahashi},
  {Tanab{\'e}}, {Takeuchi}, {Takita}, {Thomson}, {Uemizu}, {Ueno}, {Usui},
  {Verdugo}, {Wada}, {Wang}, {Watabe}, {Watarai}, {White}, {Yamamura},
  {Yamauchi}, \& {Yasuda}}]{murakami2007}
{Murakami}, H., {et~al.} 2007, \pasj, 59, 369

\bibitem[{{Negrello} {et~al.}(2009){Negrello}, {Serjeant}, {Pearson}, {Takagi},
  {Efstathiou}, {Goto}, {Burgarella}, {Jeong}, {Im}, {Lee}, {Matsuhara},
  {Oyabu}, {Wada}, \& {White}}]{negrello2009}
{Negrello}, M., {et~al.} 2009, \mnras, 394, 375

\bibitem[{{Ohyama} {et~al.}(2007){Ohyama}, {Onaka}, {Matsuhara}, {Wada}, {Kim},
  {Fujishiro}, {Uemizu}, {Sakon}, {Cohen}, {Ishigaki}, {Ishihara}, {Ita},
  {Kataza}, {Matsumoto}, {Murakami}, {Oyabu}, {Tanab{\'e}}, {Takagi}, {Ueno},
  {Usui}, {Watarai}, {Pearson}, {Takeyama}, {Yamamuro}, \&
  {Ikeda}}]{ohyama2007}
{Ohyama}, Y., {et~al.} 2007, \pasj, 59, 411

\bibitem[{{Onaka} {et~al.}(2007){Onaka}, {Matsuhara}, {Wada}, {Fujishiro},
  {Fujiwara}, {Ishigaki}, {Ishihara}, {Ita}, {Kataza}, {Kim}, {Matsumoto},
  {Murakami}, {Ohyama}, {Oyabu}, {Sakon}, {Tanab{\'e}}, {Takagi}, {Uemizu},
  {Ueno}, {Usui}, {Watarai}, {Cohen}, {Enya}, {Ootsubo}, {Pearson}, {Takeyama},
  {Yamamuro}, \& {Ikeda}}]{onaka2007}
{Onaka}, T., {et~al.} 2007, \pasj, 59, 401

\bibitem[{{Papovich} {et~al.}(2001){Papovich}, {Dickinson}, \&
  {Ferguson}}]{papovich2001}
{Papovich}, C., {Dickinson}, M., \& {Ferguson}, H.~C. 2001, \apj, 559, 620

\bibitem[{{Polletta} {et~al.}(2007){Polletta}, {Tajer}, {Maraschi},
  {Trinchieri}, {Lonsdale}, {Chiappetti}, {Andreon}, {Pierre}, {Le F{\`e}vre},
  {Zamorani}, {Maccagni}, {Garcet}, {Surdej}, {Franceschini}, {Alloin},
  {Shupe}, {Surace}, {Fang}, {Rowan-Robinson}, {Smith}, \&
  {Tresse}}]{Polletta2007}
{Polletta}, M., {et~al.} 2007, \apj, 663, 81

\bibitem[{{Rieke} {et~al.}(2004){Rieke}, {Young}, {Engelbracht}, {Kelly},
  {Low}, {Haller}, {Beeman}, {Gordon}, {Stansberry}, {Misselt}, {Cadien},
  {Morrison}, {Rivlis}, {Latter}, {Noriega-Crespo}, {Padgett}, {Stapelfeldt},
  {Hines}, {Egami}, {Muzerolle}, {Alonso-Herrero}, {Blaylock}, {Dole}, {Hinz},
  {Le Floc'h}, {Papovich}, {P{\'e}rez-Gonz{\'a}lez}, {Smith}, {Su}, {Bennett},
  {Frayer}, {Henderson}, {Lu}, {Masci}, {Pesenson}, {Rebull}, {Rho}, {Keene},
  {Stolovy}, {Wachter}, {Wheaton}, {Werner}, \& {Richards}}]{rieke2004}
{Rieke}, G.~H., {et~al.} 2004, \apjs, 154, 25

\bibitem[{{Rix} {et~al.}(2004){Rix}, {Barden}, {Beckwith}, {Bell}, {Borch},
  {Caldwell}, {H{\"a}ussler}, {Jahnke}, {Jogee}, {McIntosh}, {Meisenheimer},
  {Peng}, {Sanchez}, {Somerville}, {Wisotzki}, \& {Wolf}}]{rix2004}
{Rix}, H.-W., {et~al.} 2004, \apjs, 152, 163

\bibitem[{{Rowan-Robinson} {et~al.}(2008){Rowan-Robinson}, {Babbedge},
  {Oliver}, {Trichas}, {Berta}, {Lonsdale}, {Smith}, {Shupe}, {Surace},
  {Arnouts}, {Ilbert}, {Le F{\'e}vre}, {Afonso-Luis}, {Perez-Fournon},
  {Hatziminaoglou}, {Polletta}, {Farrah}, \& {Vaccari}}]{rowanrobinson2008}
{Rowan-Robinson}, M., {et~al.} 2008, \mnras, 386, 697

\bibitem[{{Sajina} {et~al.}(2005){Sajina}, {Lacy}, \& {Scott}}]{sajina2005}
{Sajina}, A., {Lacy}, M., \& {Scott}, D. 2005, \apj, 621, 256

\bibitem[{{Salvato} {et~al.}(2009){Salvato}, {Hasinger}, {Ilbert}, {Zamorani},
  {Brusa}, {Scoville}, {Rau}, {Capak}, {Arnouts}, {Aussel}, {Bolzonella},
  {Buongiorno}, {Cappelluti}, {Caputi}, {Civano}, {Cook}, {Elvis}, {Gilli},
  {Jahnke}, {Kartaltepe}, {Impey}, {Lamareille}, {LeFloch}, {Lilly},
  {Mainieri}, {McCarthy}, {McCracken}, {Mignoli}, {Mobasher}, {Murayama},
  {Sasaki}, {Sanders}, {Schiminovich}, {Shioya}, {Shopbell}, {Silverman},
  {Smol{\v c}i{\'c}}, {Surace}, {Taniguchi}, {Thompson}, {Trump}, {Urry}, \&
  {Zamojski}}]{salvato2009}
{Salvato}, M., {et~al.} 2009, \apj, 690, 1250

\bibitem[{{Scoville} {et~al.}(2007){Scoville}, {Aussel}, {Brusa}, {Capak},
  {Carollo}, {Elvis}, {Giavalisco}, {Guzzo}, {Hasinger}, {Impey}, {Kneib},
  {LeFevre}, {Lilly}, {Mobasher}, {Renzini}, {Rich}, {Sanders}, {Schinnerer},
  {Schminovich}, {Shopbell}, {Taniguchi}, \& {Tyson}}]{scoville2007}
{Scoville}, N., {et~al.} 2007, \apjs, 172, 1

\bibitem[{{Shapley} {et~al.}(2005){Shapley}, {Steidel}, {Erb}, {Reddy},
  {Adelberger}, {Pettini}, {Barmby}, \& {Huang}}]{shapley2005}
{Shapley}, A.~E., {Steidel}, C.~C., {Erb}, D.~K., {Reddy}, N.~A., {Adelberger},
  K.~L., {Pettini}, M., {Barmby}, P., \& {Huang}, J. 2005, \apj, 626, 698

\bibitem[{{Stansberry} {et~al.}(2007){Stansberry}, {Gordon}, {Bhattacharya},
  {Engelbracht}, {Rieke}, {Marleau}, {Fadda}, {Frayer}, {Noriega-Crespo},
  {Wachter}, {Young}, {M{\"u}ller}, {Kelly}, {Blaylock}, {Henderson},
  {Neugebauer}, {Beeman}, \& {Haller}}]{stansberry2007}
{Stansberry}, J.~A., {et~al.} 2007, \pasp, 119, 1038

\bibitem[{{Stern} {et~al.}(2005){Stern}, {Eisenhardt}, {Gorjian}, {Kochanek},
  {Caldwell}, {Eisenstein}, {Brodwin}, {Brown}, {Cool}, {Dey}, {Green},
  {Jannuzi}, {Murray}, {Pahre}, \& {Willner}}]{stern2005}
{Stern}, D., {et~al.} 2005, \apj, 631, 163

\bibitem[{{Surace} {et~al.}(2005){Surace}, {Shupe}, {Fang}, {Evans}, {Alexov},
  {Frayer}, {Lonsdale}, \& {SWIRE Team}}]{surace2005}
{Surace}, J.~A., {Shupe}, D.~L., {Fang}, F., {Evans}, T., {Alexov}, A.,
  {Frayer}, D., {Lonsdale}, C.~J., \& {SWIRE Team}. 2005, in Bulletin of the
  American Astronomical Society, Vol.~37, Bulletin of the American Astronomical
  Society, 1246--+

\bibitem[{{Surace} {et~al.}(2007){Surace}, {Krick}, {Ashby}, {Egami}, {Frayer},
  {Gorjian}, {Hora}, {Hund}, {Lacy}, {Moustakas}, {Thompson}, \&
  {Yan}}]{surace2007}
{Surace}, J.~A., {et~al.} 2007, in Bulletin of the American Astronomical
  Society, Vol.~38, Bulletin of the American Astronomical Society, 93--+

\bibitem[{{Werner} {et~al.}(2004){Werner}, {Roellig}, {Low}, {Rieke}, {Rieke},
  {Hoffmann}, {Young}, {Houck}, {Brandl}, {Fazio}, {Hora}, {Gehrz}, {Helou},
  {Soifer}, {Stauffer}, {Keene}, {Eisenhardt}, {Gallagher}, {Gautier}, {Irace},
  {Lawrence}, {Simmons}, {Van Cleve}, {Jura}, {Wright}, \&
  {Cruikshank}}]{werner2004}
{Werner}, M.~W., {et~al.} 2004, \apjs, 154, 1

\bibitem[{{Williams} {et~al.}(1996){Williams}, {Blacker}, {Dickinson}, {Dixon},
  {Ferguson}, {Fruchter}, {Giavalisco}, {Gilliland}, {Heyer}, {Katsanis},
  {Levay}, {Lucas}, {McElroy}, {Petro}, {Postman}, {Adorf}, \&
  {Hook}}]{williams1996}
{Williams}, R.~E., {et~al.} 1996, \aj, 112, 1335

\bibitem[{{Wilson} {et~al.}(2003){Wilson}, {Eikenberry}, {Henderson},
  {Hayward}, {Carson}, {Pirger}, {Barry}, {Brandl}, {Houck}, {Fitzgerald}, \&
  {Stolberg}}]{wilson2003}
{Wilson}, J.~C., {et~al.} 2003, in Society of Photo-Optical Instrumentation
  Engineers (SPIE) Conference Series, Vol. 4841, Society of Photo-Optical
  Instrumentation Engineers (SPIE) Conference Series, ed. M.~{Iye} \& A.~F.~M.
  {Moorwood}, 451--458

\end{thebibliography}

\end{document}